\documentclass[twocolumn]{aastex701}
\usepackage{amsmath}
\usepackage[T1]{fontenc}
\usepackage{hyperref}

\begin{document}

\title{Electromagnetic Signatures of Supermassive Binary Black Holes: Synchrotron, Self-Lensing Flares, and Jet Precession}

\author[orcid=0000-0003-0292-2773]{Hong-Xuan Jiang}
\email{hongxuan\_jiang@sjtu.edu.cn}
\affiliation{Tsung-Dao Lee Institute, Shanghai Jiao Tong University, Shengrong Road 520, Shanghai, 201210, China}

\author[0000-0003-0750-3543]{Xinyu Li}
\email{xinyuli@tsinghua.edu.cn}
\affiliation{Department of Astronomy, Tsinghua University, 30 Shuangqing Rd, Beijing, 100084, China}

\author[orcid=0009-0004-8669-2411]{Jing-Ze Xia}
\email{jackxia@sjtu.edu.cn}
\affiliation{Tsung-Dao Lee Institute, Shanghai Jiao Tong University, Shengrong Road 520, Shanghai, 201210, China}

\author[orcid=0000-0002-8131-6730]{Yosuke Mizuno}
\email{mizuno@sjtu.edu.cn}
\affiliation{Tsung-Dao Lee Institute, Shanghai Jiao Tong University, Shengrong Road 520, Shanghai, 201210, China}
\affiliation{School of Physics and Astronomy, Shanghai Jiao Tong University, 
800 Dongchuan Road, Shanghai, 200240, China}
\affiliation{Key Laboratory for Particle Physics, Astrophysics and Cosmology, Shanghai Key Laboratory for Particle Physics and Cosmology, Shanghai Jiao-Tong University, 800 Dongchuan Road, Shanghai, 200240, China}
\affiliation{Institut f\"ur Theoretische Physik, Goethe-Universit\"at Frankfurt, Max-von-Laue-Stra{\ss}e 1, D-60438 Frankfurt am Main, Germany}

\author{Ziri Younsi}
\email{z.younsi@ucl.ac.uk}
\affiliation{Mullard Space Science Laboratory, University College London, Holmbury St. Mary, Dorking, Surrey, RH5 6NT, UK}

\author{Christian M. Fromm}
\email{christian.fromm@uni-wuerzburg.de}
\affiliation{Institut f\"ur Theoretische Physik und Astrophysik, Universit\"at W\"urzburg, Emil-Fischer-Str. 31, D-97074 W\"urzburg, Germany}
\affiliation{Institut f\"ur Theoretische Physik, Goethe-Universit\"at Frankfurt, Max-von-Laue-Stra{\ss}e 1, D-60438 Frankfurt am Main, Germany}

\correspondingauthor{Hong-Xuan Jiang}
\email{hongxuan\_jiang@sjtu.edu.cn}
\correspondingauthor{Xinyu Li} 
\email{xinyuli@tsinghua.edu.cn}
\correspondingauthor{Yosuke Mizuno}
\email{mizuno@sjtu.edu.cn}

\begin{abstract}
The recent evidence for a nanohertz gravitational wave background from Pulsar Timing Arrays highlights the urgent need to identify electromagnetic counterparts to supermassive binary black holes. Here, we perform global 3D general relativistic magnetohydrodynamic (GRMHD) simulations of a secondary black hole (mass ratio $q=0.1$) interacting with a Magnetically Arrested Disk around a primary black hole using a time-dependent superposed Kerr-Schild metric and post-processed general relativistic radiation transfer calculations based on thermal electron distribution function (eDF). We explore three orbital configurations: a vertical impact orbit, a coplanar embedded orbit, and a high-spin, eccentric, inclined scenario. Despite clear orbital periodicity and recurrent shock formation, the thermal synchrotron light curves frequently lack expected shock-induced flares. In vertical impacts, shock brightenings are typically sub-dominant to the stochastic MAD variability of the primary black hole, unless viewed at specific alignment phases. Conversely, coplanar orbits produce distinctive, rapid flares driven by gravitational self-lensing. We identify a frequency-dependent emission hierarchy: the primary black hole dominates sub-millimeter flux, while the secondary dominates near-infrared emission due to higher electron temperatures in thermal eDF. Finally, spin-orbit coupling drives Lense-Thirring precession, yielding twisted, wobbling jets that following the tilt and precession of the primary BH. Crucially, we show that intrinsic MAD turbulence can easily mask shock-induced radio flares, making self-lensing flares a more reliable electromagnetic counterpart candidate for supermassive binary black holes.
\end{abstract}


\section{Introduction}
supermassive binary black holes (SMBBHs) are a natural outcome of hierarchical galaxy growth and frequent mergers \citep{1980Natur.287..307B,2010A&ARv..18..279V}. 
Pulsar Timing Arrays (PTAs) are expected to be most sensitive to the nanohertz stochastic gravitational-wave background from the cosmological SMBBH population \citep{2019A&ARv..27....5B,2025arXiv250500797K}, and recent PTA data sets report compelling evidence for such a signal \citep{2023ApJ...951L...8A,2026arXiv260109481A,2023A&A...678A..48E,2023A&A...678A..49E,2023A&A...678A..50E}. 
Electromagnetic identification of SMBBHs is therefore increasingly important, both to interpret PTA detections and to enable multimessenger science with future space-based missions such as the LISA \citep{2017arXiv170200786A,2019BAAS...51g..77T}. 

In the electromagnetic domain, persistent (quasi-)periodic outbursts and jet modulations remain widely used diagnostics for SMBBH candidates, with the blazar OJ~287 as a well-known, but physically distinct geometrically thin-disk/blazar example \citep{1988ApJ...325..628S,1996A&A...315L..13S,1996ApJ...460..207L,2008Natur.452..851V,2016ApJ...819L..37V,2018ApJ...866...11D,2020ApJ...894L...1L,xia2026magneticconfigurationimprintsquasiperiodic}. 
Recent polarimetric VLBI observations reveal a compact, twisted, magnetized inner jet and spatially resolved polarization variability in the innermost jet base \citep{2022ApJ...932...72Z,2026A&A...705A..23G}. 
In addition, nearly edge-on SMBBHs can produce short-duration self-lensing flares when one BH magnifies emission from the other \citep[e.g.,][]{2018MNRAS.474.2975D,2020MNRAS.495.4061H}; such events are expected to be sharply peaked and approximately time-symmetric, providing a clean signature against stochastic accretion variability \citep{2022PhRvD.105j3010D,2022PhRvL.128s1101D,2024PhRvD.109j3014K}. 

Physically interpreting these signatures requires strong-field GRMHD modeling of accretion and jet launching near both BHs. 
Standard Newtonian circumbinary disk simulations capture large-scale gas dynamics but miss key relativistic effects near the horizons \citep[e.g.,][]{2024ApJ...970..156D,2020ApJ...889..114M}. 
Full GRMHD simulations that evolve the plasma in a time-dependent binary spacetime are computationally expensive for long-duration, high-resolution studies \citep[e.g.,][]{2012PhRvL.109v1102F,2015CQGra..32q5009E,2024PhRvD.109j3024F,2025PhRvD.112d3004M}. 
A practical approach is to evolve GRMHD on an approximate, time-dependent binary black hole (BBH) spacetime: post-Newtonian (PN) near-zone methods enabled early studies but required excising the cavity near the horizons \citep{2012ApJ...755...51N,2021PhRvD.104d4041C}, while a superposed Kerr--Schild construction enables long-duration, horizon-resolving simulations and shows good agreement with full numerical relativity \citep{2021ApJ...913...16L,2024arXiv240313308C}. 
Recent applications have begun to explore the impacts on magnetically arrested disks (MADs) and separate thin-disk crossing models motivated by systems such as OJ~287, finding signatures of enhanced accretion, outflows, and orbital precession \citep{2021ApJ...917...43S, 2024ApJ...967...70R,2025ApJ...979L..24R,2025ApJ...993L..22R}. 

MADs, in which magnetic flux accumulates near the BH and powers efficient Blandford--Znajek jets \citep{1977MNRAS.179..433B,2011MNRAS.418L..79T}, can reproduce key observational features of M~87$^\ast$ including variability and polarization structure \citep{2012MNRAS.426.3241N,2015MNRAS.447...49S,2019ApJ...875L...6E,2019ApJ...875L...5E,2021ApJ...910L..12E,2025A&A...704A..91E}. 
However, the observable electromagnetic signatures of a secondary BH perturbing such an M~87$^\ast$-like low-luminosity active galactic nucleus (LLAGN) system remain poorly quantified. 
We therefore use M~87$^\ast$ as a fiducial target to study a broader class of LLAGN hosting MAD-like, radiatively inefficient accretion flows, and to quantify how a secondary BH would perturb their multi-wavelength emission. 
In this work, we address this by performing controlled global 3D GRMHD simulations of a MAD torus perturbed by a secondary BH with mass ratio $q=0.1$ on a time-dependent approximate BBH metric, and connect the dynamics to observables via general relativistic radiative transfer (GRRT) post-processing across radio to near-infrared (NIR) wavelengths. By moving beyond parameterized analytical models, these first-principles calculations reveal how intrinsic turbulent noise masks binary interaction signatures, establishing a rigorous predictive baseline for targeting SMBBHs with next-generation facilities like the ngEHT and GRAVITY+.

\begin{table*}
\centering
\caption{Run suite. All binary runs adopt $q=0.1$ and an initial separation $d=30\,r_{\rm g}$. Inclination is measured relative to the initial torus midplane.
All production runs use the same Cartesian domain, $[-1024,1024]^3\,r_{\rm g}$, a $40^3$ root grid with $40^3$ cells per meshblock, static mesh refinement (SMR) to level 8, and BH-tracking adaptive mesh refinement (AMR) to level 10, giving $\Delta x_{\rm min}=0.05\,r_{\rm g}$ in the finest cells. The columns $N_{H,1}$ and $N_{H,2}$ give the number of finest cells across the event-horizon diameter, $N_{H,i}\equiv 2r_{H,i}/\Delta x_{\rm min}$ with $r_{H,i}=M_i[1+(1-a_i^2)^{1/2}]$, where $a_i$ is the dimensionless spin listed in the table.}
Run labels: $\tt{VT}$ = vertical impact, $\tt{CP}$ = coplanar embedded, $\tt{EP}$ = eccentric precessing (high-spin), and $\tt{BASE}$ = single-BH baseline.
\label{tab:sim_parameters}
\begin{tabular}{lccccccc}
\hline \hline
run & $a_1$ & $a_2$ & Inclination ($i$) & Eccentricity ($e$) & $N_{H,1}$ & $N_{H,2}$ & Description \\
\hline
$\tt{VT}$   & 0      & 0.9375  & $90^\circ$ & 0   & $80$ & $5.4$ & Vertical impact (transient shock) \\
$\tt{CP}$   & 0      & 0.9375  & $0^\circ$  & 0   & $80$ & $5.4$ & Coplanar (persistently embedded) \\
$\tt{EP}$   & 0.9375 & 0.9375  & $90^\circ$ & 0.3 & $54$ & $5.4$ & High-spin, eccentric, precessing \\
$\tt{BASE}$ & 0      & \textemdash & \textemdash & \textemdash & $80$ & \textemdash & Single BH \\
\hline
\end{tabular}
\end{table*}

\section{Numerical Method}
\label{sec:numerical}

We perform global 3D GRMHD simulation runs of a MAD torus perturbed by a secondary BH in a time-dependent binary spacetime. We adopt geometric units with $G=c=M_1=1$, where $M_1$ is the primary BH mass. The secondary mass is fixed to $M_2=qM_1=0.1$, and we define $r_{\rm g}\equiv GM_1/c^2$ as the gravitational radius of the primary. Throughout the paper, times written in units of ${\rm M}$ use this simulation mass unit, i.e., ${\rm M}\equiv M_1$.
Table~\ref{tab:sim_parameters} summarizes the run suite; all binary runs adopt $q=0.1$ and are initialized with separation $d=30\,r_{\rm g}$.

Following \cite{2024ApJ...967...70R}, we fix the primary BH at the coordinate origin and allow its spin axis, along with the corresponding jet and disk orientations, to precess due to spin-orbit coupling.
We consider three binary configurations (runs~$\tt{VT}$, $\tt{CP}$, and $\tt{EP}$) and a single-BH baseline case (run~$\tt{BASE}$).
Hereafter we refer to these configurations as runs~$\tt{VT}$, $\tt{CP}$, $\tt{EP}$, and $\tt{BASE}$ (Table~\ref{tab:sim_parameters}).
Runs~$\tt{VT}$ and $\tt{CP}$ use a non-spinning primary ($a_1=0$) as controlled experiments: run~$\tt{VT}$ is a vertical impact orbit ($i=90^\circ$) and run~$\tt{CP}$ is a coplanar embedded orbit ($i=0^\circ$).
Run~$\tt{EP}$ adopts high spins ($a_1=a_2=0.9375$) and an eccentric ($e=0.3$), initially $90^\circ$ tilted secondary-BH orbit, producing a strongly time-dependent, precessing configuration.
To ensure a smooth dynamical transition, the simulation initially evolves a single primary BH. The secondary BH is then gradually introduced into the system, with its mass smoothly increasing from zero to $M_2=0.1M_1$ during the interval $t = 10{,}000\,\rm M$ to $11{,}000\,\rm M$. This mass ramp is a numerical regularization rather than a physical mass-growth model. It avoids an impulsive metric, excision, and floor adjustment when the secondary BH is introduced, and all quantitative diagnostics used for the light curves and morphology analysis are measured well after the ramp interval. The GRMHD implementation, including the PN orbit, superposed Kerr--Schild metric, and mesh setup, is described in the following subsections.

\subsection{Binary orbit and spin evolution}
\label{app:pn}

The dynamics of the BBH system are governed by equations of motion derived from the PN framework, incorporating both conservative and radiation-reaction terms as described in \citet{2014LRR....17....2B, 2015JHEP...09..219L}, with:
\begin{equation}
    \dot{\mathbf{r}} = \mathbf{v}, \quad \dot{\mathbf{S}}_A = \boldsymbol{\Omega}_A \times \mathbf{S}_A, \quad A \in \{1, 2\}\,,
\end{equation}
where $\mathbf{r}$ is the position vector, $\mathbf{v}$ is the velocity, and $\mathbf{S}_A$ denotes the spin vector of the $A$-th black hole. We define the following kinematic variables required for the equations of motion:
\begin{equation}
    r = |\mathbf{r}|, \quad \mathbf{n} = \frac{\mathbf{r}}{r}, \quad v^2 = \mathbf{v} \cdot \mathbf{v}, \quad \dot{r} = \mathbf{n} \cdot \mathbf{v}.
\end{equation}

The acceleration is decomposed into four terms representing the non-spinning (ns) and spin-dependent contributions, further split into conservative (cons) and radiation-reaction (RR) effects:
\begin{equation}
    \mathbf{a} = \mathbf{a}_{\rm ns}^{\rm cons} + \mathbf{a}_{\rm ns}^{\rm RR} + \mathbf{a}_{\rm spin}^{\rm cons} + \mathbf{a}_{\rm spin}^{\rm RR}.
\end{equation}
These terms are expressed via PN expansions:
\begin{align}
    \mathbf{a}_{\rm ns}^{\rm cons} &= \mathbf{a}_{0{\rm PN}} + \mathbf{a}_{1{\rm PN}} + \mathbf{a}_{2{\rm PN}}, \\
    \mathbf{a}_{\rm ns}^{\rm RR} &= \mathbf{a}_{2.5{\rm PN}}^{\rm RR} + \mathbf{a}_{3.5{\rm PN}}^{\rm RR} + \mathbf{a}_{4{\rm PN}}^{\rm RR}, \\
    \mathbf{a}_{\rm spin}^{\rm cons} &= \mathbf{a}_{\rm SO}^{1.5{\rm PN}} + \mathbf{a}_{\rm SO}^{2.5{\rm PN}} + \mathbf{a}_{\rm SS}^{2{\rm PN}} + \mathbf{a}_{\rm SS}^{3{\rm PN}}, \\
    \mathbf{a}_{\rm spin}^{\rm RR} &= \mathbf{a}_{\rm SO}^{2.5{\rm PN,RR}} + \mathbf{a}_{\rm SO}^{3.5{\rm PN,RR}}.
\end{align}
Here, $\mathbf{a}_{\rm spin}^{\rm cons}$ and $\mathbf{a}_{\rm spin}^{\rm RR}$ account for spin-orbit (SO) and spin-spin (SS) interactions. The spin-precession terms include leading-order (LO), next-to-leading-order (NLO), and quadrupole-monopole (QM) contributions, with frequency $\boldsymbol{\Omega}_A$ given by:
\begin{equation}
    \boldsymbol{\Omega}_A = \boldsymbol{\Omega}_{A, \rm SO}^{\rm NLO} + \boldsymbol{\Omega}_{A, \rm SS}^{\rm LO} + \boldsymbol{\Omega}_{A, \rm SS}^{\rm NLO} + \boldsymbol{\Omega}_{A, \rm QM},
\end{equation}
which encompasses geodetic and LT precessions, spin-spin interactions, and quadrupole-monopole self-spin coupling.

The initial tangential velocity $v_\phi$ for quasi-circular orbits is determined by solving the radial force balance equation using only the conservative terms:
\begin{equation}
    a_r^{\rm cons} + \frac{v_\phi^2}{r} = 0.
\end{equation}
The full PN system is then integrated forward in time to evolve the binary orbit and spins.

\subsection{Superposed Kerr--Schild binary metric}
\label{app:sks}

We adopt the superposed Kerr--Schild construction \citep{2021ApJ...913...16L, 2024arXiv240313308C}, in which the covariant metric is expressed as a Minkowski background plus Kerr--Schild perturbations from two boosted Kerr BHs,
\begin{equation}
g_{\mu\nu}
=
\eta_{\mu\nu}
+
\sum_{A=1}^{2} H_A\,l^{(A)}_\mu l^{(A)}_\nu,
\end{equation}
where $H_A$ is the Kerr--Schild scalar for BH $A$ and $l^{(A)}_\mu$ is the corresponding ingoing null covector. Orbital motion is incorporated by boosting each BH's Kerr--Schild coordinate from its instantaneous comoving frame to the simulation frame,
\begin{equation}
l^{(A)}_\mu
=
(\Lambda_A)_\mu{}^\alpha\,l^{(A,{\rm rest})}_\alpha,
\end{equation}
with $\Lambda_A$ the Lorentz boost determined by the instantaneous BH velocity from the PN trajectory.

\subsection{GRMHD setup on a time-dependent superposed Kerr--Schild spacetime}
\label{app:grmhd}

\begin{figure*}[!t]
    \centering
    \includegraphics[width=0.85\linewidth]{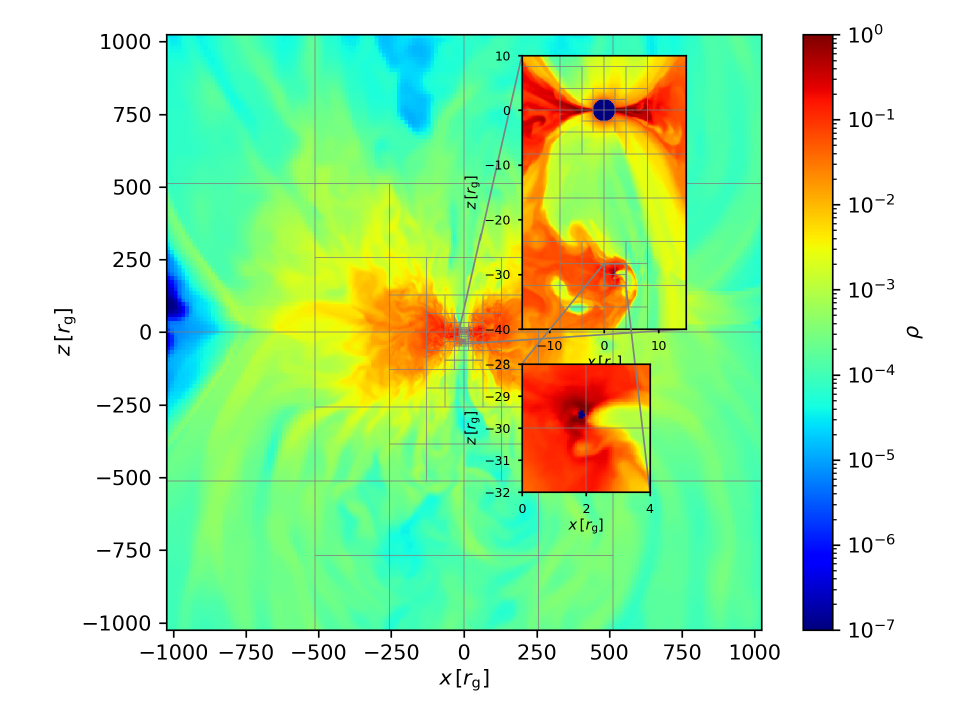}
    \caption{Gas density slice from run~$\tt{VT}$ at $t=26{,}410\,\rm M$ (edge-on slice), overlaid with the meshblock boundaries.}
    \label{fig:meshblock}
\end{figure*}
Following \citet{2021ApJ...913...16L, 2024ApJ...967...70R}, the background spacetime is modeled using the superposed Kerr--Schild metric, with the BBH trajectory and spin evolution supplied by high-order PN evolution \citep{2014LRR....17....2B, 2015JHEP...09..219L}, where the metric and orbit prescriptions are described in Sections~\ref{app:sks} and~\ref{app:pn}, respectively. The GRMHD equations are solved with the \texttt{DynGRMHD} module in \texttt{AthenaK} \citep{Stone2024}, adopting an ideal-gas equation of state with $\Gamma=5/3$.

To capture both the global MAD accretion flow and the horizon-scale dynamics around the secondary BH, we employ SMR together with AMR, using a BH-tracking criterion to maintain uniform resolution around each horizon. All models use the mesh setup summarized in Table~\ref{tab:sim_parameters}, with a Cartesian domain $[-1024,1024]^3\,r_{\rm g}$, a root grid of $40^3$ cells, $40^3$ cells per meshblock, SMR to level 8, and BH-tracking AMR to level 10. The finest cells have $\Delta x_{\rm min}=0.05\,r_{\rm g}$, corresponding to an effective uniform resolution of $40960^3$ over the full domain. The event-horizon diameter is resolved by $80$ finest cells for the non-spinning-primary runs and by $54$ finest cells for the high-spin primary in run~$\tt{EP}$; the high-spin secondary horizon diameter is resolved by $5.4$ finest cells. An example density slice from run~$\tt{VT}$ illustrating the refinement layout and the boundary of meshblocks is shown in Fig.~\ref{fig:meshblock}. 
A resolution sensitivity test for the secondary BH horizon, based on a refined restart, is presented in Appendix~\ref{app:secondary_resolution}. Additional validation tests of the single-BH GRMHD evolution, the $q=0$ metric limit, and the GRRT metric implementation are presented in Appendices~\ref{app:single_bh_validation} and~\ref{app:grrt_metric_validation}.

In GRMHD simulations, we set the magnetization ceiling $\sigma_{\rm ceiling}=100$. We initialize the flow with a Fishbone--Moncrief hydrostatic equilibrium torus \citep{1976ApJ...207..962F} with parameters of $r_{\rm min}=20\,r_{\rm g}$ and $r_{\rm max}=40\,r_{\rm g}$, with radii measured in the primary-based length unit. The torus is embedded with a poloidal magnetic field to reach the MAD regime. The magnetic field is initialized as
\begin{equation}
\begin{aligned}
    A_{\rm \phi}\propto &(\rho - 0.01)(r/r_{\rm in})^3\sin^3\theta \exp{(-r/400)}\\
    & \sin(2\pi(r-r_{\rm in})),
\end{aligned}
\end{equation}
with the field strength set by the minimum plasma $\beta_{\rm min} = 100$, where $\beta$ is the ratio of gas pressure ($p_{\rm gas}$) to magnetic pressure ($p_{\rm mag}$).

\subsection{GRRT equations, electron thermodynamics, and post-processing setup}
\label{app:grrt_details}

The GRRT calculations are performed with \texttt{BHOSS} \citep{2012A&A...545A..13Y,2020IAUS..342....9Y} using the fast-light approximation. We solve the GRRT equation along null geodesics,
\begin{equation}
\frac{d\mathbf{S}}{d\lambda}=\mathbf{j}-\mathbf{K}\mathbf{S},
\end{equation}
where $\mathbf{S}=(I,Q,U,V)^{\top}$ is the Stokes vector, $\mathbf{j}$ is the emissivity vector, and the matrix $\mathbf{K}$ contains the plasma absorption and Faraday rotation and conversion coefficients. Electron temperatures are specified using the $R$-$\beta$ prescription \citep{2016A&A...586A..38M},
\begin{equation}
\frac{T_i}{T_e}
=
R_{\rm high}\frac{\beta^2}{1+\beta^2}
+
R_{\rm low}\frac{1}{1+\beta^2},
\end{equation}
where $\beta$ is the plasma beta and $(R_{\rm high},R_{\rm low})=(10,1)$ in our fiducial models, motivated by R--$\beta$ fits to two-temperature GRMHD simulations \citep{2015MNRAS.454.1848R,2023MNRAS.522.2307J,2024A&A...688A..82J, 2026A&A...707A..27X, 2021MNRAS.506..741M} and by recent comparisons showing that R--$\beta$ prescriptions remain a useful approximation to the electron-temperature scale in two-temperature MAD models \citep{2025ApJ...981..145M, 2021MNRAS.506..741M}. Here $R_{\rm high}$ controls the ion-to-electron temperature ratio in high-$\beta$ disk gas, while $R_{\rm low}$ controls the ratio in low-$\beta$ magnetized regions, yielding relatively cool electrons in the dense disk and hot electrons in the magnetized funnel/outflow. Two-temperature calculations provide valuable guidance for the expected temperature ratios. However, their electron-heating closures are themselves uncertain and are often motivated by particular dissipation channels, such as turbulence or reconnection, which need not apply uniformly throughout a MAD flow. We therefore adopt the simpler R--$\beta$ prescription as a controlled parametrization that limits dependence on uncertain electron-heating subgrid physics and isolates the radiative signatures of the binary dynamics.

We compute synthetic observables by post-processing GRMHD snapshots with \texttt{BHOSS}, modified to ray-trace on the superposed Kerr--Schild spacetime in Cartesian coordinates. The observables are computed for an edge-on viewing inclination $\theta_{\rm obs}=90^\circ$, and we consider two azimuthal viewing angles, $\phi_{\rm obs}=0^\circ$ and $90^\circ$, to isolate cases with strong and minimal effects from the line of sight (LOS), respectively. For the GRRT calculations, we adopt M~87$^\ast$ as a fiducial target, with physical primary mass scale $M_{1,\rm phys}=6.5\times10^9\,{\rm M_\odot}$ and source distance $D=16.8\,{\rm Mpc}$. We exclude highly magnetized regions by applying a cutoff at $\sigma>5$. For the main light-curve and image calculations, we assume a thermal Maxwell--J\"uttner eDF. In Section~\ref{sec:nonthermal_effects}, we also perform an exploratory nonthermal power-law electron injection to the GRRT calculations for runs~$\tt{VT}$ and~$\tt{CP}$. We retain the thermal component, but inject a power-law tail with fixed slope $p=2.5$ within the region $r_2<10\,r_{\rm g}$, where $r_2$ is the distance from the secondary BH. We use the shock diagnostic $\chi_{\rm shock}$ defined in Section~\ref{sec:shock_3D} as a proxy for the local shock strength, assigning a nonthermal energy fraction 
\begin{equation}
    \epsilon_{\rm nth} =
  \begin{cases}
  0, & \chi_{\rm shock}\le 0.1,\\
  \dfrac{0.5}{0.9}(\chi_{\rm shock}-0.1), & 0.1<\chi_{\rm shock}<1,\\
  0.5, & \chi_{\rm shock}\ge 1.
  \end{cases}
\end{equation}
Here $\epsilon_{\rm nth}$ is the fraction of the local electron internal energy assigned to the nonthermal power-law component. The details of the implementation of the nonthermal synchrotron calculation follow \citet{2021MNRAS.507.5281C}. The maximum Lorentz factor is set to $\gamma_{\max}=10^6$, which is insensitive to our results.
The weakly compressive regions with $\chi_{\rm shock}\le 0.1$ remain purely thermal, while regions with $\chi_{\rm shock}\ge 1$ saturate at $\epsilon_{\rm nth}=0.5$. We also include a phenomenological synchrotron-cooling cutoff, motivated by nonthermal flare models in MAD flows \citep{2022MNRAS.511.3536S, 2025ApJ...990...81J}, by suppressing the nonthermal synchrotron coefficients above $\nu_{\rm br}=2.5\times10^{15}(100\,{\rm G}/|B|)\,{\rm Hz}$ with a factor $\exp(-\nu/\nu_{\rm br})$. The nonthermal runs use the same $R$--$\beta$ prescription, $\sigma>5$ cutoff, and density normalization as their corresponding thermal calculations, including $\rho_{\rm unit}=1.5\times10^{-20}\,{\rm g\,cm^{-3}}$ for the $\tt{VT}$ and $\tt{CP}$ comparisons. A representative synthetic image check of this nonthermal prescription is presented in Appendix~\ref{app:nonthermal_image_validation}.

We set the density normalization of the GRRT calculations by matching the single-BH baseline (run~$\tt{BASE}$) to a total flux of $\simeq 0.5\,{\rm Jy}$ at $230\,{\rm GHz}$ for M~87$^\ast$. For runs~$\tt{VT}$, $\tt{CP}$, and $\tt{BASE}$, we adopt $\rho_{\rm unit}=1.5\times10^{-20}\,{\rm g\,cm^{-3}}$, calibrated using snapshots from the quasi-steady interval $t=25{,}000$--$30{,}000\,\rm M$. For run~$\tt{EP}$, we use a lower normalization, $\rho_{\rm unit}=2\times10^{-21}\,{\rm g\,cm^{-3}}$, to obtain a comparable total $230\,{\rm GHz}$ flux. We focus on $230\,{\rm GHz}$ because it is the standard EHT observing band and anchors the M~87$^\ast$ flux normalization, on $86\,{\rm GHz}$ to trace larger-scale jet morphology accessible to mm-VLBI, and on $138\,{\rm THz}$ as a representative NIR band where high temperature electron emission and lensing contrast are stronger. For imaging, we use a field of view of $80\,r_{\rm g}\times 80\,r_{\rm g}$ and a fixed camera resolution of $1{,}000\times 1{,}000$ pixels for all \texttt{BHOSS} images, which is sufficient to resolve the photon ring of the secondary BH.

This GRRT setup is used for all synthetic light curves and images below unless explicitly stated otherwise.

\section{Results}

\begin{figure*}[!t]
\centering 	
\includegraphics[width=\linewidth]{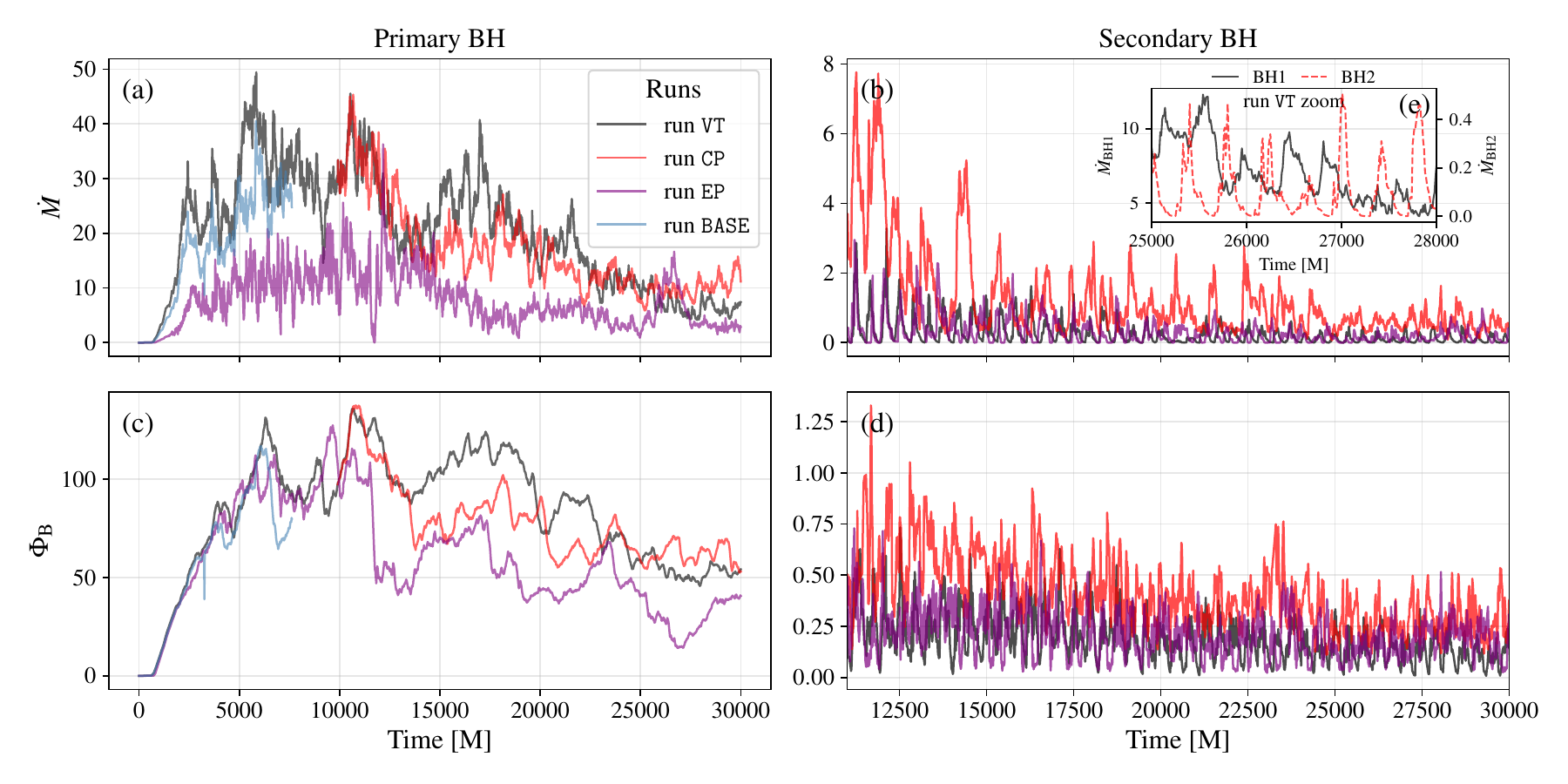}
\caption{Mass accretion rates $\dot{M}$ and magnetic fluxes $\Phi_{\rm B}$ measured at the event horizons of the primary and secondary BHs. Subscripts BH1 and BH2 denote the primary and secondary BHs, respectively.
(a) Accretion rate of the primary BH, $\dot{M}_{\rm BH1}$, for all runs (black: run~$\tt{VT}$; red: run~$\tt{CP}$; purple: run~$\tt{EP}$; blue: run~$\tt{BASE}$).
(b) Accretion rate of the secondary BH, $\dot{M}_{\rm BH2}$, for runs~$\tt{VT}$, $\tt{CP}$, and $\tt{EP}$ (colors as in panel a).
(c) Horizon-threading magnetic flux of the primary BH, $\Phi_{\rm B,BH1}$, for all runs.
(d) Horizon-threading magnetic flux of the secondary BH, $\Phi_{\rm B,BH2}$, for runs~$\tt{VT}$, $\tt{CP}$, and $\tt{EP}$.
In all runs, the secondary BH is ramped in gradually over $t=10{,}000$--$11{,}000\,\rm M$; consequently, runs~$\tt{VT}$, $\tt{CP}$, and $\tt{BASE}$ are identical for $t<10{,}000\,\rm M$, and their divergences occur only after $t\gtrsim 11{,}000\,\rm M$.
(e) Close-up view of the accretion rate evolution for run~$\tt{VT}$ during $t \in [2.5, 2.8] \times 10^{4}\,\rm M$, comparing $\dot{M}_{\rm BH1}$ (black solid; left axis) and $\dot{M}_{\rm BH2}$ (red dashed; right axis).}
\label{fig:mdot_run12}
\end{figure*}

\begin{figure}[!t]
\centering 	
\includegraphics[width=\linewidth]{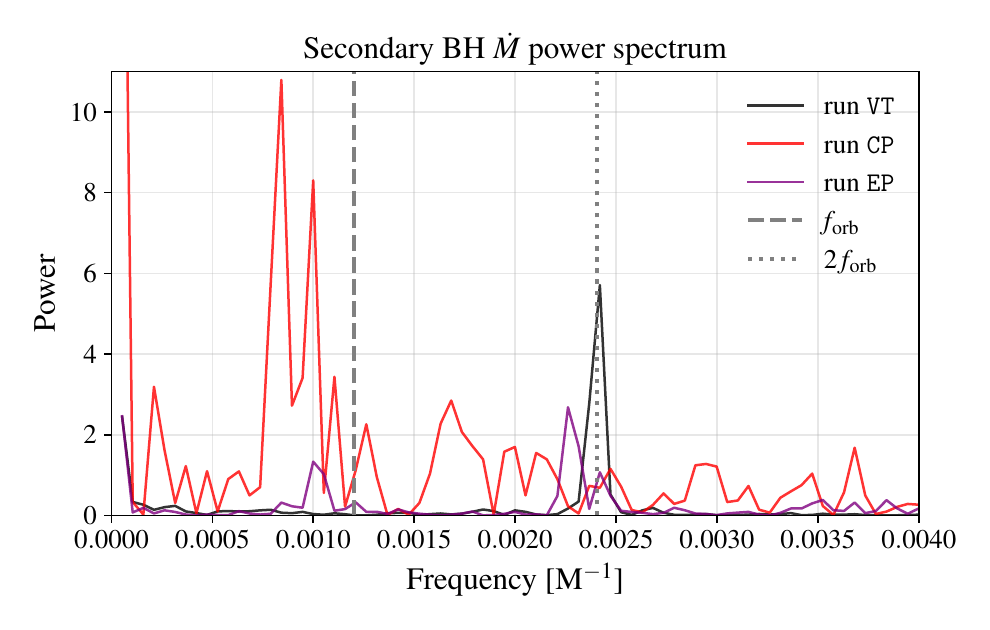}
\caption{Power spectrum of the secondary-BH accretion rate, $\dot{M}_{\rm BH2}$. Solid curves show run~$\tt{VT}$ (black), run~$\tt{CP}$ (red), and run~$\tt{EP}$ (purple).
Vertical gray dashed and dotted lines mark $f_{\rm orb}$ and $2f_{\rm orb}$, respectively.
}
\label{fig:bh2_spectrum}
\end{figure}

\subsection{GRMHD processes during BH--disk interaction} \label{sec:BH_disk_interaction}

In runs~$\tt{VT}$ and $\tt{CP}$, we consider a Schwarzschild BH as primary, under which the secondary BH remains on an approximately circular and fixed orbit. In Fig.~\ref{fig:mdot_run12}, we plot the accretion rate and magnetic flux measured on horizon surfaces around both BHs.
For each BH, we construct the measurement surface in its local Kerr--Schild coordinates, 
accounting for the Lorentz boost due to orbital motion. The mass accretion rate 
is computed as
\begin{equation}
    \dot{M} = -\int_{S} \rho\, u^r \sqrt{-g}\, d\theta\, d\phi,
\end{equation}
where $\rho$ is the rest-mass density, $u^r$ is the radial component of the 
4-velocity (with the gradient $\partial r/\partial x^i$ properly boosted to 
the lab frame), and $\sqrt{-g}\,d\theta\,d\phi$ is the invariant surface 
measure. The magnetic flux through the horizon is calculated by:
\begin{equation}
    \Phi_{\rm B} = \frac{1}{2} \int_{S} \left| b^r u^0 - b^0 u^r \right| 
    \sqrt{-g}\, d\theta\, d\phi,
\end{equation}
where $b^\mu$ is the 4-magnetic field in the fluid frame, related to the 
coordinate magnetic field $B^i$ by $b^0 = u_i B^i$ and 
$b^i = (B^i + b^0 u^i)/u^0$, which corresponds to the magnitude of 
the dual Faraday tensor component $|{}^*F^{rt}|$. 

As shown in Fig.~\ref{fig:mdot_run12}a, compared to the single-BH run~$\tt{BASE}$, run~$\tt{VT}$ does not exhibit a clear long-term deviation in $\dot{M}_{\rm BH1}$ after the secondary BH is introduced. However, we can see a clear modulation of the primary BH accretion rate associated with the vertical passages of the secondary BH through the torus. In the zoomed-in panel (e), where we plot $\dot{M}_{\rm BH1}$ and $\dot{M}_{\rm BH2}$, each time the secondary BH crosses the torus, the accretion rate of the primary BH shows a small bump followed by a decrease; the next passage produces a similar rise in $\dot{M}_{\rm BH1}$, repeating this pattern in each impact.
In contrast, run~$\tt{CP}$ shows an initially elevated $\dot{M}_{\rm BH1}$ following the introduction of the secondary BH, but this enhancement diminishes later and results in a slightly lower $\dot{M}_{\rm BH1}$ after $t=15{,}000\,\rm M$. When the secondary BH is first introduced, it interacts with the torus in a manner similar to the torus--BH impacts in run~$\tt{VT}$.
It drives shocks which compress the gas and the funnel material toward the primary BH, temporarily increasing $\dot{M}_{\rm BH1}$. However, over time, the secondary BH evacuates gas along its orbit, reducing the available supply for accretion. This depletion is analogous to the low-density cavity seen in circumbinary disk simulations, both in earlier hydrodynamic work and in recent MHD/MAD studies \citep{2019ApJ...871...84M,2020ApJ...889..114M,2024ApJ...970..156D,2024ApJ...973L..19M,2025PhRvD.111h1304M,2025arXiv250816855W}.
The difference is most pronounced in run~$\tt{EP}$. With a rapidly spinning primary BH, both $\dot{M}_{\rm BH1}$ and $\Phi_{\rm B,BH1}$ are lower than in Schwarzschild cases (runs~$\tt{VT}$, $\tt{CP}$, and $\tt{BASE}$).

In Fig.~\ref{fig:mdot_run12}b and d, the secondary-BH diagnostics $\dot{M}_{\rm BH2}$ and $\Phi_{\rm B,BH2}$ exhibit strongly periodic bursts in run~$\tt{VT}$ (black solid curves) that coincide with the secondary BH penetrating the torus. By contrast, run~$\tt{CP}$ (red solid curves) displays a more irregular, stochastic evolution, consistent with turbulence driven by magnetorotational instability. 
Although run~$\tt{EP}$ is initialized with a $90^\circ$ inclined secondary-BH orbit, the orbital plane undergoes substantial reorientation over time (Section~\ref{sec:orientation}), alternating between more vertical and more coplanar phases. As a result, enhancements in $\dot{M}_{\rm BH2}$ are less sharply periodic than in run~$\tt{VT}$, yet remain more coherent than the fully stochastic behavior in run~$\tt{CP}$.

Fig.~\ref{fig:bh2_spectrum} shows the power spectrum of $\dot{M}_{\rm BH2}$ for the three binary runs.
Both run~$\tt{VT}$ and run~$\tt{CP}$ exhibit periodic variability, yet display distinct characteristic frequencies.
In run~$\tt{VT}$, where the secondary BH impacts the torus twice per orbit, the power spectrum shows a prominent peak aligned precisely with $2\times f_{\rm orb}$.
In contrast, run~$\tt{CP}$ reveals a characteristic frequency slightly lower than the orbital frequency, $f_{\rm orb}$.
Theoretically, the orbital motion of a secondary BH embedded in a perfectly axisymmetric flow should not induce quasi-periodic oscillations.
However, MAD flows inherently develop non-axisymmetric structures (e.g., rotating magnetic flux bundles).
Since these structures co-rotate with the flow, the secondary BH must ``catch up'' to the rotating pattern during each orbit, i.e., $f_{\rm obs} \approx f_{\rm orb} - f_{\rm flow}$.
Consequently, the effective interaction frequency is shifted to a value slightly lower than $f_{\rm orb}$. 
In run~$\tt{EP}$, the power spectrum shows two clear peaks near $f_{\rm orb}$ and $2f_{\rm orb}$, reflecting the fact that the secondary BH alternates between phases that resemble embedded, coplanar motion (favoring a once-per-orbit modulation) and phases with more vertical disk crossings (favoring a twice-per-orbit impact signature). This mixed behavior is consistent with the time-dependent reorientation of the orbit seen in Section~\ref{sec:orientation}, which naturally imprints both harmonics rather than a single dominant frequency.

\subsection{Shock and mini-jet in GRMHD simulation}
\label{sec:shock_3D}

\begin{figure*}[!t]
\centering 	
\includegraphics[height=.7\textheight]{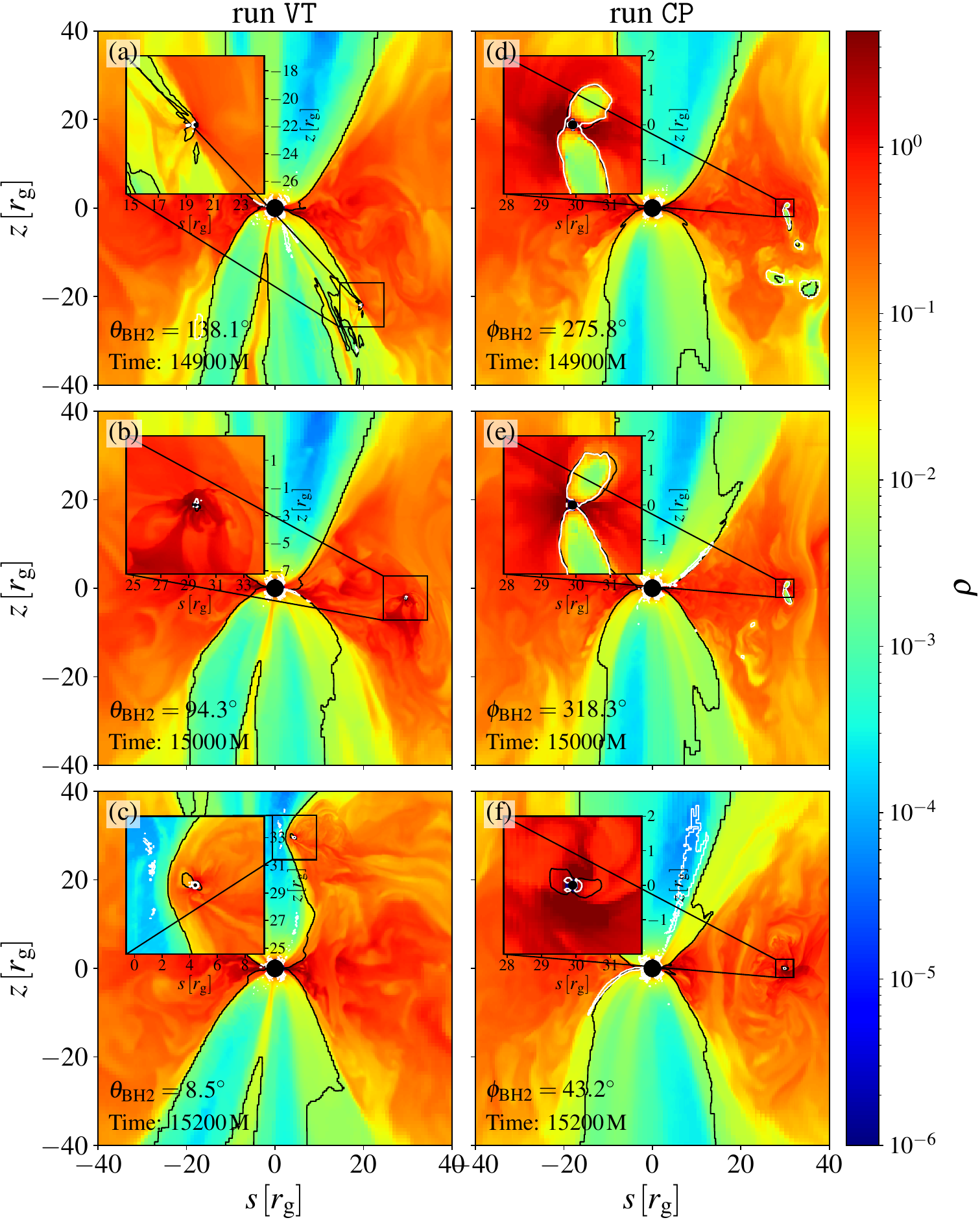}
\caption{Evolution of the gas density ($\rho$) in run~$\tt{VT}$ (left) and run~$\tt{CP}$ (right) across three temporal snapshots: $t=14{,}900\,\rm M$, $15{,}000\,\rm M$, and $15{,}200\,\rm M$. To capture the orbital dynamics, each panel displays a vertical ($s, z$) slice that dynamically tracks the secondary BH. The horizontal coordinate $s$ represents the radial distance along the instantaneous separation vector connecting the two masses, with the primary anchored at the origin ($s=0$), while $z$ denotes the vertical height in units of $r_{\rm g}$. The logarithmic color map visualizes the density field, overlaid with black and blue solid contours that delineate the magnetization $\sigma=1$ and Bernoulli parameter $hu_t=-1.1$ boundary, respectively. Magnified inset panels highlight the immediate localized environment of the secondary. Each frame is annotated with the simulation time and the secondary's current orbital phase angle ($\theta_{\rm BH2}$ for run~$\tt{VT}$; $\phi_{\rm BH2}$ for run~$\tt{CP}$).}
\label{fig:2D_slice}
\end{figure*}

\begin{figure*}[!t]
\centering 	
\includegraphics[height=.4\linewidth]{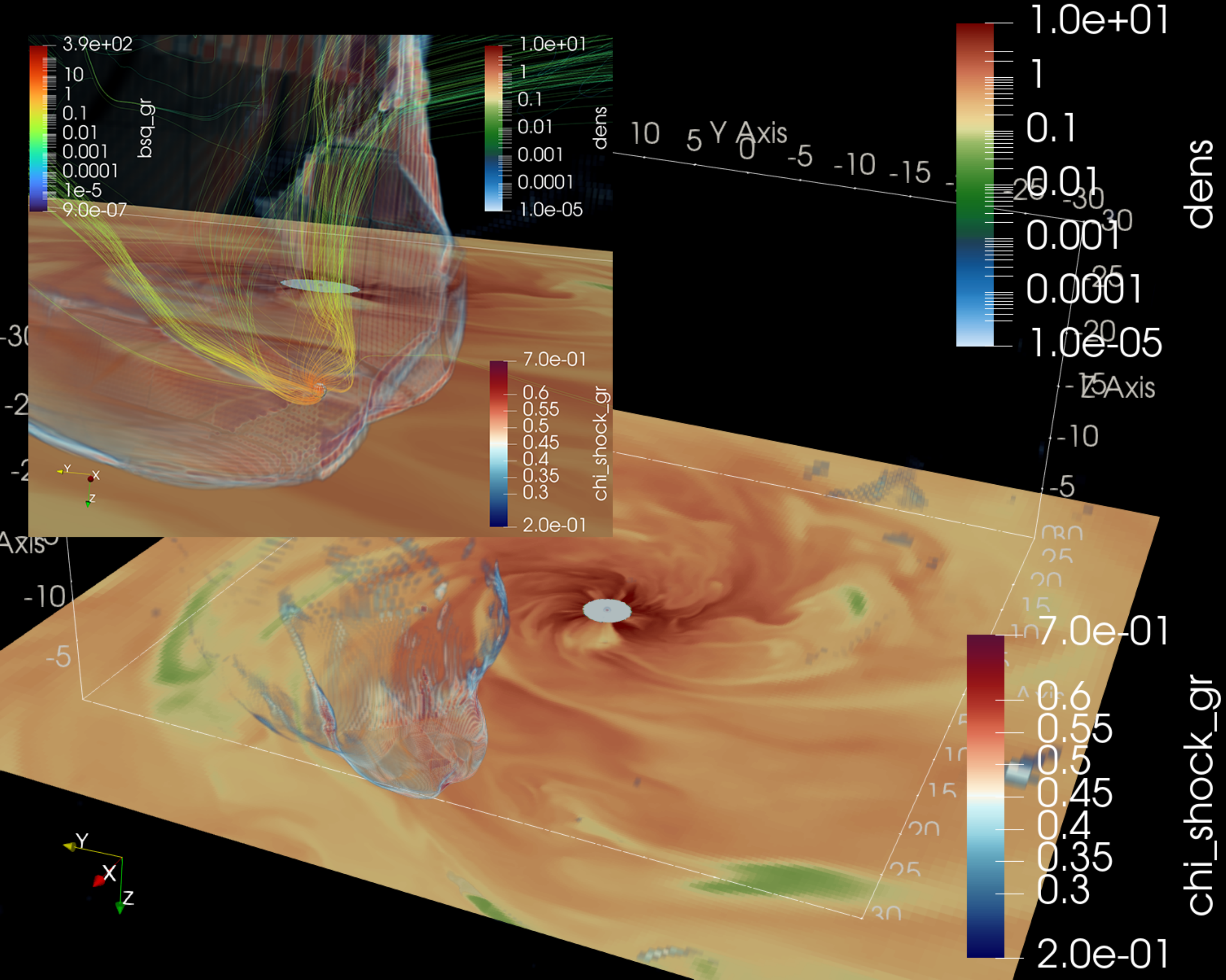}\hfill
\includegraphics[height=.4\linewidth]{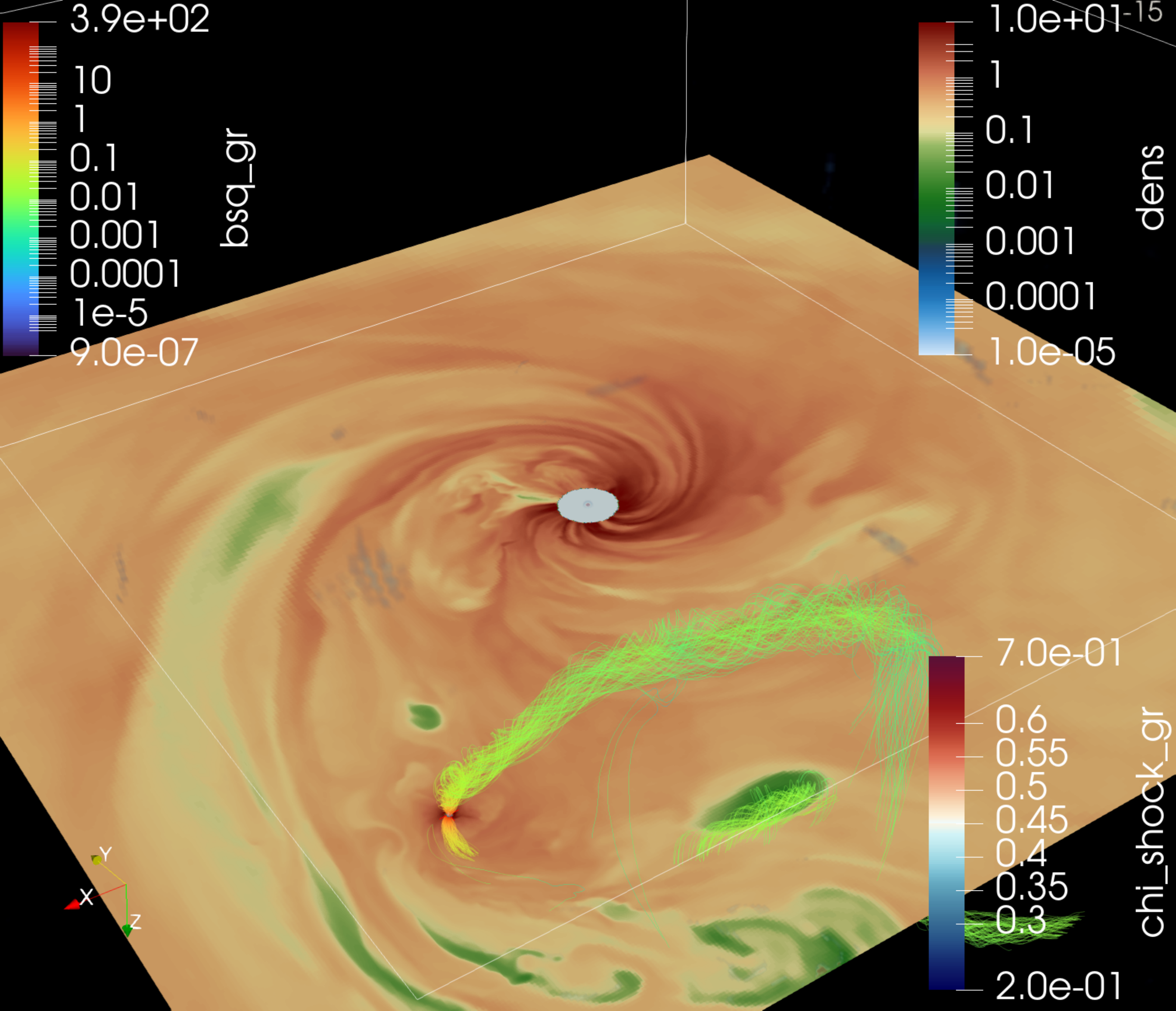}
\caption{3D visualization of the shock diagnostic and plasma structure near the secondary BH.
Each panel shows a density slice together with volume rendering of the shock indicator $\chi_{\rm shock}$.
Left: run~$\tt{VT}$ snapshot; right: run~$\tt{CP}$ snapshot.
Insets and color bars indicate the plotted density and $\chi_{\rm shock}$ ranges (and the displayed field-line overlay where shown).}
\label{fig:shock_bfield}
\end{figure*}

To understand the shock structures and transient jet-like outflows that develop in our simulations, we first compare the large-scale flow morphology in runs $\tt{VT}$ and $\tt{CP}$.
In Fig.~\ref{fig:2D_slice}, we show a time sequence of vertically sliced density maps for run~$\tt{VT}$ (left column) and run~$\tt{CP}$ (right column). For each snapshot, the slice is taken in the instantaneous plane spanned by the BH--BH separation vector and the global $z$-axis. We rotate the slicing plane every time so that its horizontal coordinate $s$ always lies along the current line connecting the primary and secondary BHs. In this co-rotating slice, the primary BH remains fixed at $s=0$, while the secondary appears at positive $s$ at its instantaneous orbital phase (annotated in each panel by $\theta_{\rm BH2}$ for run~$\tt{VT}$ and $\phi_{\rm BH2}$ for run~$\tt{CP}$). The colors show the density ($\rho$) on a logarithmic scale, and the black contour marks the magnetization $\sigma=1$, which serves as a practical jet boundary (for either the primary jet or any outflow powered by the secondary). The contours of $hu_t=-1.1$ indicate that these regions are gravitationally unbound and overlap with the highly magnetized region identified as the mini-jet. The run~$\tt{VT}$ sequence captures a transient disk penetration: as the secondary plunges through and then emerges from the dense torus, it drives strong compressions, while simultaneously entraining a compact overdense cloud that remains bound briefly before being stripped and swept away once it enters the primary BH's jet. By contrast, in run~$\tt{CP}$ the secondary stays embedded in the torus throughout this interval, maintaining a steady local gas supply. Consistent with this sustained feeding, density slices reveal episodic jet activity from the secondary BH: a collimated feature is visible at intermediate times (panels d--e), but it weakens and is absent by $t=15{,}200\,\rm M$ (panel f), indicating a quenched mini-jet state at that epoch.  

Figure~\ref{fig:shock_bfield} provides a complementary three-dimensional view of the BH--disk interaction by visualizing the compressive structures and magnetic lines in the vicinity of the secondary BH. We introduce a dimensionless shock diagnostic,
\begin{equation}
\chi_{\rm shock} \equiv \frac{\max\left(-\nabla_i v^i,0\right)}{\ell_{\rm proper}/c_{\rm fast}},
\end{equation}
where $\nabla_i v^i$ is the covariant velocity divergence, $\ell_{\rm proper}$ is the proper local cell length scale, and $c_{\rm fast}$ is the local fast magnetosonic speed. This quantity compares the compression time to the fast-mode crossing time: $\chi_{\rm shock}\ll 1$ corresponds to weak compression, while $\chi_{\rm shock}\gtrsim 1$ flags regions where compression is rapid enough to generate shock-like structures.

In the run~$\tt{VT}$ snapshot (Fig.~\ref{fig:shock_bfield}a), $\chi_{\rm shock}$ shows a compact, high-compression region concentrated around the secondary, coincident with the strongest disturbance of the surrounding gas. The accompanying zoom-in, with magnetic field lines overlaid, shows pronounced bending and wrapping of the field in the compressed layer, consistent with flux-freezing as the secondary drives strong vertical interaction with the torus material. In the run~$\tt{CP}$ snapshot (Fig.~\ref{fig:shock_bfield}b), the same dynamic range shows little visible high-$\chi_{\rm shock}$ volume, indicating that the flow near the secondary is comparatively less compressive. In this case, the secondary largely co-moves with the local disk gas, reducing relative velocities and suppressing the formation of a strong shock. Despite the weaker shocks, the more persistent inflow onto the rapidly spinning secondary BH ($a_2=0.9375$) supports intermittent mini-jet launching. As binary orbits, the resulting outflow is twisted and deflected by the ambient torus and the global field/jet environment (seen in the magnetic lines in Fig.~\ref{fig:shock_bfield}b).

\subsection{\texorpdfstring{Observational signatures of runs~$\tt{VT}$ and $\tt{CP}$}{Observational signatures of runs VT and CP}}
\label{sec:obs_signature}

\begin{figure*}[!t]
\centering 	
\includegraphics[width=\linewidth]{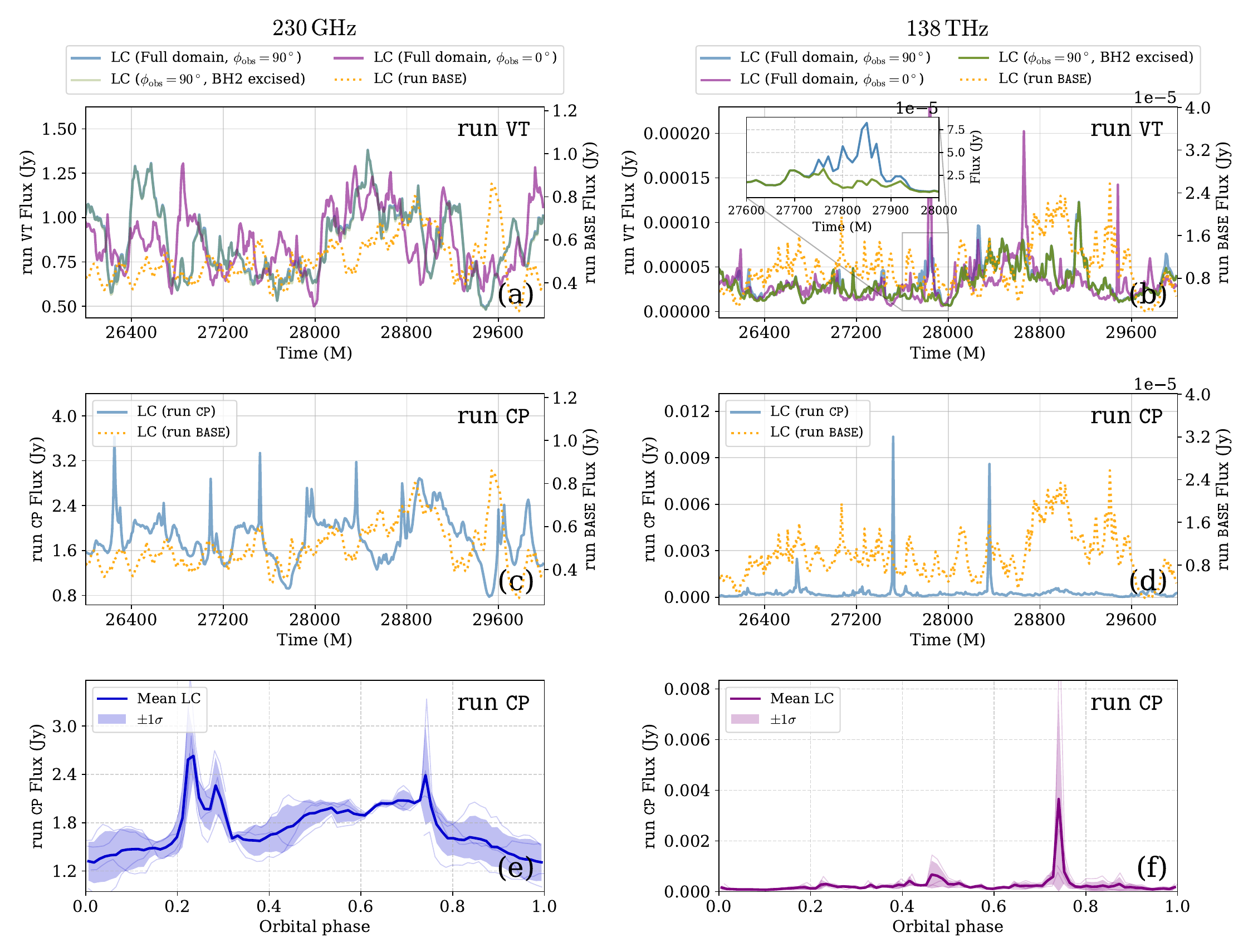}
\caption{
Thermal synchrotron light curves from GRRT post-processing. Panels (a)--(d) show $t=26{,}000$--$30{,}000\,\rm M$ time series, with the binary-run flux on the left axis and the single-BH control run~$\tt{BASE}$ on the right axis (orange dotted). The top row shows run~$\tt{VT}$, and the middle row shows run~$\tt{CP}$.
(a) run~$\tt{VT}$ at $230\,\mathrm{GHz}$ for $\phi_{\rm obs}=90^\circ$ (blue) and $\phi_{\rm obs}=0^\circ$ (purple), and for $\phi_{\rm obs}=90^\circ$ with plasma within $r_2<10\,r_{\rm g}$ excised (green, lower opacity).
(b) Same as (a) at $138\,\mathrm{THz}$; the inset ($t=27{,}600$--$28{,}000\,\rm M$) highlights a particularly strong shock-driven flare during which the full-domain flux is $\sim3\times$ higher than the excised case, whereas differences are otherwise modest.
(c) run~$\tt{CP}$ at $230\,\mathrm{GHz}$ compared to run~$\tt{BASE}$.
(d) Same as (c) at $138\,\mathrm{THz}$.
(e,f) Orbit-folded run~$\tt{CP}$ light curves versus orbital phase at $230\,\mathrm{GHz}$ and $138\,\mathrm{THz}$. Faint curves show individual orbits; the thick curve and shaded band give the mean and $\pm1\sigma$.
}
\label{fig:lc_BH2}
\end{figure*}

To obtain the observational signatures of the BH--disk interaction discussed in Section~\ref{sec:BH_disk_interaction}, we perform GRRT calculations on the GRMHD snapshots.
Using the same BBH metric as the GRMHD simulation, we obtain the thermal synchrotron light curves from runs~$\tt{VT}$, $\tt{CP}$, and $\tt{BASE}$ in Fig.~\ref{fig:lc_BH2}. 

For the vertical-impact case (panels a and b in Fig.~\ref{fig:lc_BH2}), to investigate the impact of gravitational lensing, two LOS azimuths are considered: $\phi_{\rm obs} = 90^\circ$ and $\phi_{\rm obs} = 0^\circ$. 
For $\phi_{\rm obs} = 90^\circ$ (blue solid line), the LOS is perpendicular to the orbital plane of the secondary BH, resulting in no lensing effects. In contrast, for $\phi_{\rm obs} = 0^\circ$ (purple solid line), the observer and the binary BHs are aligned along the LOS twice per orbit, leading to strong gravitational lensing effects during these alignments.
Although the secondary BH exhibits highly periodic enhancements in $\dot{M}_{\rm BH2}$ (Fig.~\ref{fig:mdot_run12}b), for $\phi_{\rm obs}=90^\circ$, where the lensing effect is absent, both the $230\,\mathrm{GHz}$ and $138\,\mathrm{THz}$ light curves remain stochastic and closely resemble the turbulence-driven variability of the single-BH case (run~$\tt{BASE}$).
To quantify the shock contribution from the BH--disk interaction, we excise the plasma surrounding the secondary BH within $r_2 < 10\,r_{\rm g}$, effectively removing the major portion of the shock, as shown by the green solid lines in panels (a) and (b). 
At a specific time, for example, when $t = 27{,}840\,\rm M$, the shock induces a sharply spiked emission in the NIR light curve, which is $\sim 3$ times stronger than the emission without the shock (see the zoomed-in panel in Fig.~\ref{fig:lc_BH2}b).
However, such pronounced effects are rare. Only a few shocks are strong enough to impact the $138\,\rm THz$ light curve noticeably. For most of the simulation time, the contribution from the shock is negligible, as evidenced by the light curves overlapping closely regardless of whether the shocked region is excised.

When the LOS is aligned with the BBH separation vector, the light curves develop extremely narrow, high-contrast flares, most prominently in the NIR band. This behavior is evident in panel~b, where three sharp bursts appear on top of the underlying MAD-driven variability. The same phenomenology is also present in the coplanar run (panels~(c) and (d)), in which both the 230~GHz and NIR light curves exhibit short, spiked flares. 
In this geometry, the spikes are naturally interpreted as self-lensing events occurring near the two alignment phases per orbit: the emission from the immediate vicinity of either BH is intermittently magnified when the observer and the two BHs approach near alignment along the LOS. In particular, the strongest amplification occurs when the primary BH lies between the secondary and the observer, since it acts as a much more powerful gravitational lens and can strongly magnify the secondary's compact emission region. These brief lensing episodes produce narrow bursts superposed on the stochastic, turbulence-driven baseline emission.

\subsection{Shock-injected nonthermal emission}
\label{sec:nonthermal_effects}

\begin{figure*}[!t]
\centering
\includegraphics[width=\textwidth]{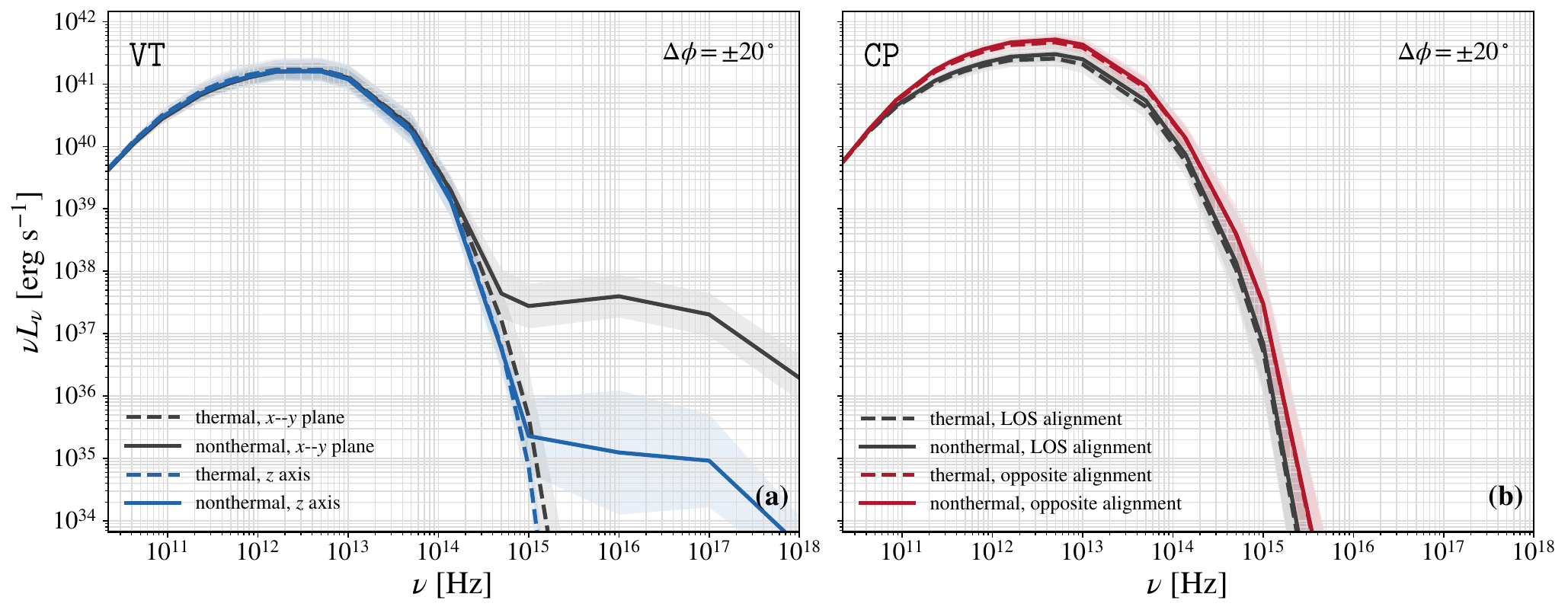}
\caption{Thermal and shock-injected nonthermal spectral energy distributions (SEDs) for runs~$\tt{VT}$ and~$\tt{CP}$. Curves show the mean SED in selected orbital-phase windows, computed in logarithmic space, and shaded regions indicate the corresponding standard-deviation scatter. Dashed curves use the thermal eDF, while solid curves include a shock-injected nonthermal eDF. Left: run~$\tt{VT}$, split between orbital phases where the secondary lies near the $x$--$y$ plane ($0^\circ,180^\circ$) and phases where it lies near the vertical axis ($90^\circ,270^\circ$). Right: run~$\tt{CP}$, split between the two LOS-alignment phases where the line of sight approximately connects the two BHs ($\phi_{\rm BH2}=0^\circ$ and $180^\circ$).}
\label{fig:nonthermal_sed_phase}
\end{figure*}

The thermal GRRT calculations above are a conservative baseline because they omit shock-accelerated electrons. To estimate the possible size of this effect, we repeat the post-processing for runs~$\tt{VT}$ and~$\tt{CP}$ with a phenomenological nonthermal eDF: a power-law tail with fixed index $p=2.5$ is added in the neighborhood of the secondary BH, and its energy fraction is related to the local shock diagnostic $\chi_{\rm shock}$. This prescription is not a kinetic acceleration model; it is intended to test where the compression identified in the GRMHD data would radiate if it efficiently injected nonthermal electrons.

Figure~\ref{fig:nonthermal_sed_phase} shows that the nonthermal correction is primarily a high-frequency effect. At sub-millimeter frequencies, the thermal and nonthermal SEDs remain close. Thus, the main conclusion from Fig.~\ref{fig:lc_BH2} still holds: primary MAD variability can hide the secondary-driven interaction at radio wavelengths. At higher frequencies $\nu\gtrsim10^{15}\,\rm Hz$, however, the shock-injected nonthermal emission can produce a much larger enhancement from the thermal model. In run~$\tt{VT}$ this behavior is physically associated with the high-$\chi_{\rm shock}$ compression around the secondary shown in Fig.~\ref{fig:shock_bfield}a. The vertical impact creates a localized acceleration site, and the resulting nonthermal electrons preferentially brighten the high-frequency SED. 

The coplanar case provides an important contrast. The right panel of Fig.~\ref{fig:shock_bfield} shows no comparable high-$\chi_{\rm shock}$ volume near the secondary, because the secondary largely co-moves with the local disk gas. Accordingly, the run~$\tt{CP}$ panel in Fig.~\ref{fig:nonthermal_sed_phase} should not be interpreted as evidence for strong shock-powered emission. Its sharp time-domain flares are still primarily produced by self-lensing during LOS alignment, as discussed above. The nonthermal calculation mainly shows that any compact high-frequency emission associated with the secondary can be magnified strongly during those alignments, even when the local shock diagnostic is weak.

Although not included in our GRRT post-process, inverse Compton scattering could further modify the high-frequency SED. As a rough estimate, for our fiducial density normalization a region with $\rho_{\rm code}\sim1$ has $n_e\simeq9.0\times10^3\,{\rm cm^{-3}}$, giving a Thomson depth $\tau_{\rm T}\simeq5.7\times10^{-6}(L/r_{\rm g})$ across a path length $L$ for the M~87$^\ast$ mass scaling. If the same region contains thermal electrons with $\Theta_e\simeq50$, the characteristic single-scattering frequency boost is $\sim16\Theta_e^2\simeq4\times10^4$. Therefore, seed photons near $\nu_s\sim10^{13}\,{\rm Hz}$ would be upscattered into the $\sim10^{15}$--$10^{18}\,{\rm Hz}$ band, which overlaps the dominant frequencies of the nonthermal emission (see Fig.~\ref{fig:nonthermal_sed_phase}). The corresponding thermal Compton parameter is $y\sim16\Theta_e^2\tau_{\rm T}\simeq0.23(L/r_{\rm g})(\rho_{\rm code}/1)$. If a fraction $f_{\rm cov}$ of the global synchrotron seed luminosity near $10^{13}\,{\rm Hz}$, $\nu L_{\nu,{\rm seed}}\sim10^{41}$--$10^{42}\,{\rm erg\,s^{-1}}$, actually passes through this hot accretion flow, then the synchrotron self-Compton (SSC) luminosity is $\nu L_{\nu,{\rm SSC}}\sim y f_{\rm cov}\nu L_{\nu,{\rm seed}}\sim2\times10^{40}$--$2\times10^{41} f_{\rm cov}(L/r_{\rm g})(\rho_{\rm code}/1)\,{\rm erg\,s^{-1}}$. This estimate is larger than the synchrotron-only high-frequency tail of run~$\tt{VT}$ in Fig.~\ref{fig:nonthermal_sed_phase} by a factor of $\sim10^2$--$10^4 f_{\rm cov}(L/r_{\rm g})(\rho_{\rm code}/1)$. Therefore, Compton scattering possibly dominates the nonthermal synchrotron peak, making it invisible in the SED.

\subsection{Synthetic horizon-scale images of the BBH system}

\begin{figure*}[!t]
\centering 	
\includegraphics[height=.33\linewidth]{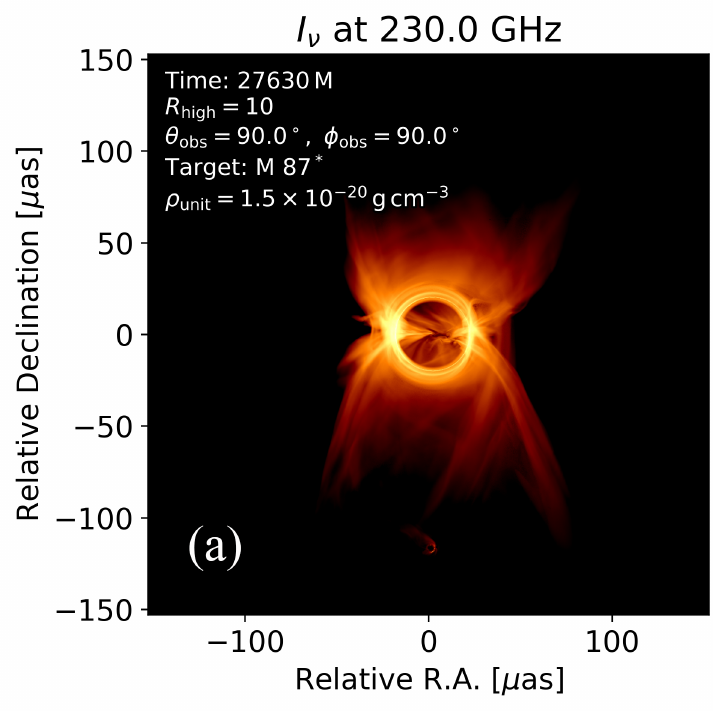}
\includegraphics[height=.33\linewidth]{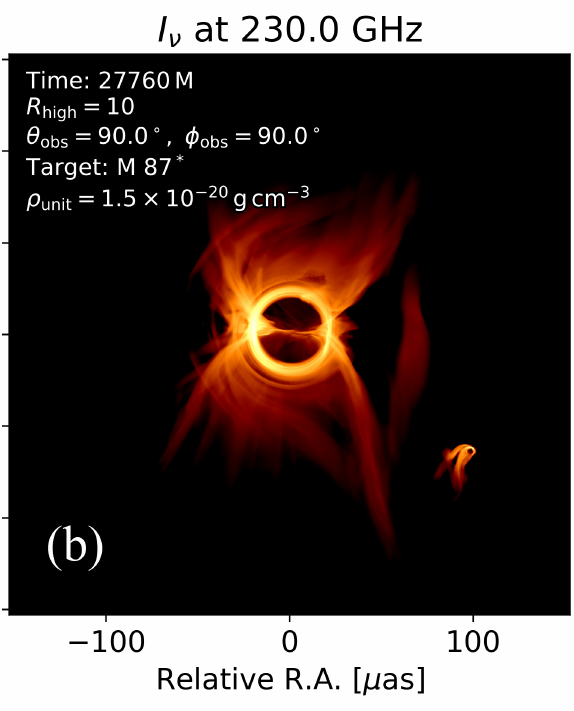}
\includegraphics[height=.33\linewidth]{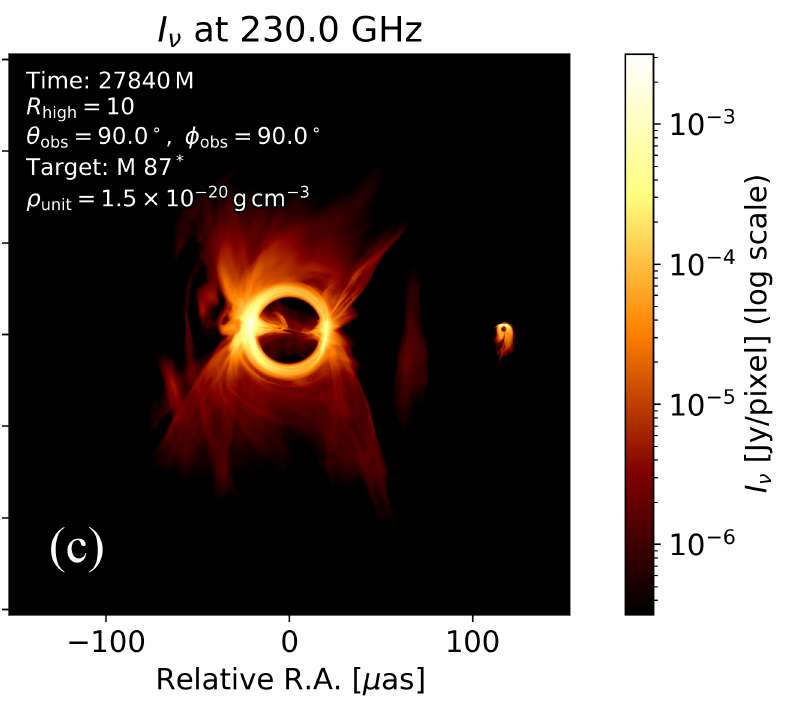}
\includegraphics[height=.33\linewidth]{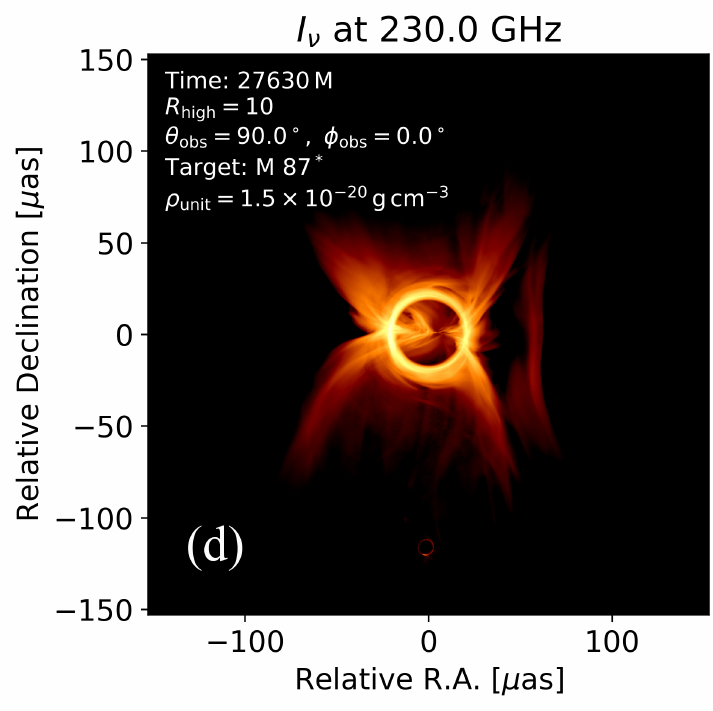}
\includegraphics[height=.33\linewidth]{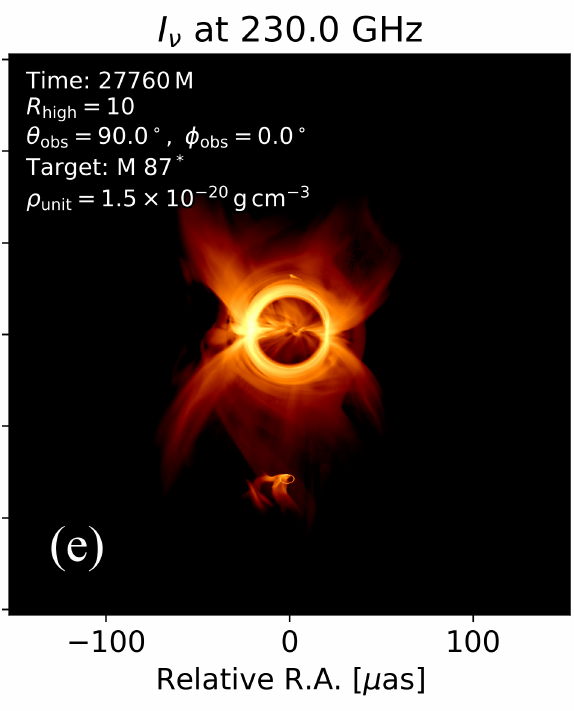}
\includegraphics[height=.33\linewidth]{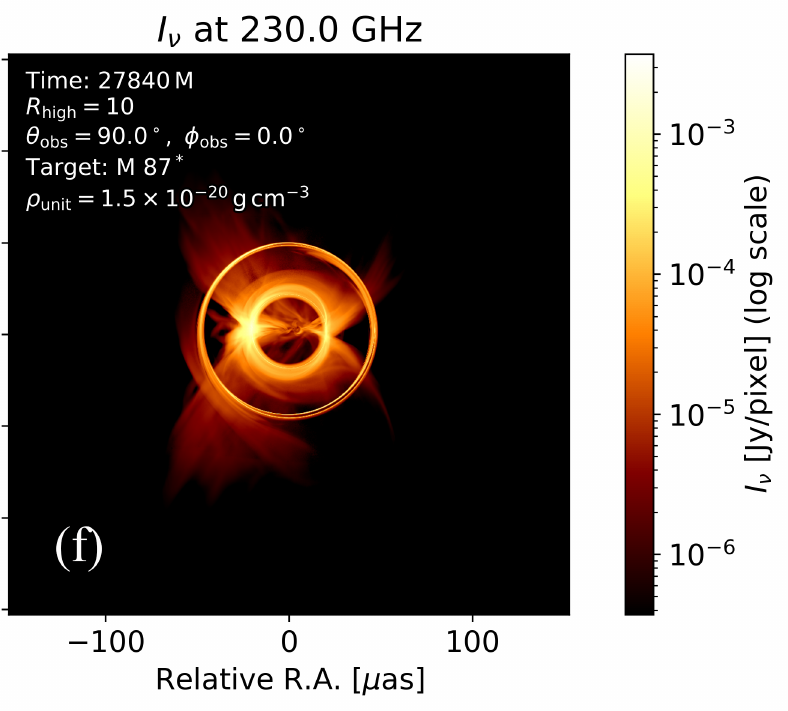}
\includegraphics[height=.33\linewidth]{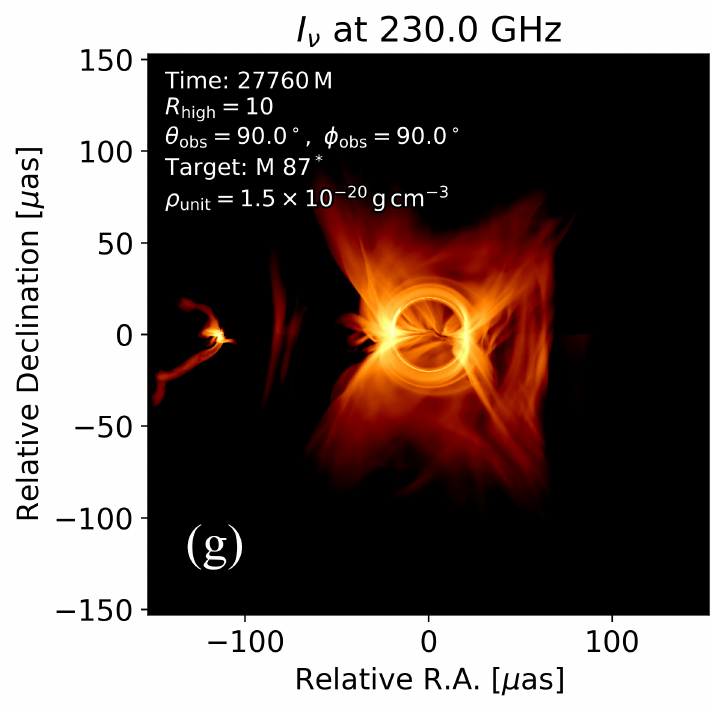}
\includegraphics[height=.33\linewidth]{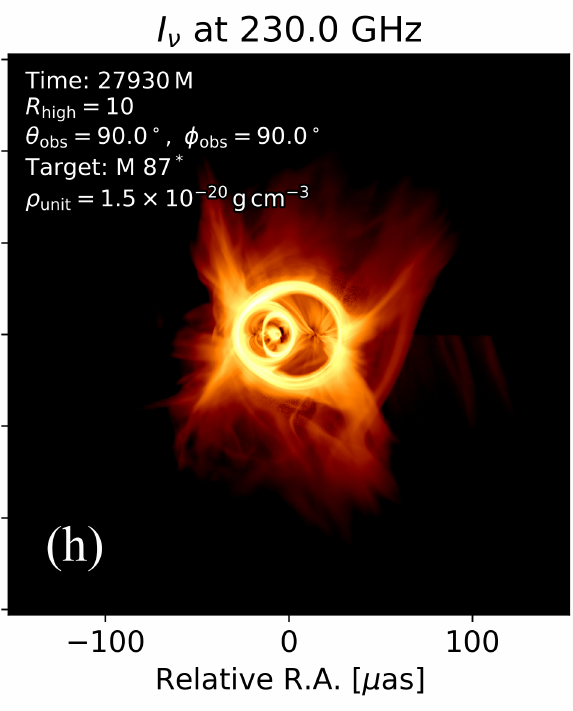}
\includegraphics[height=.33\linewidth]{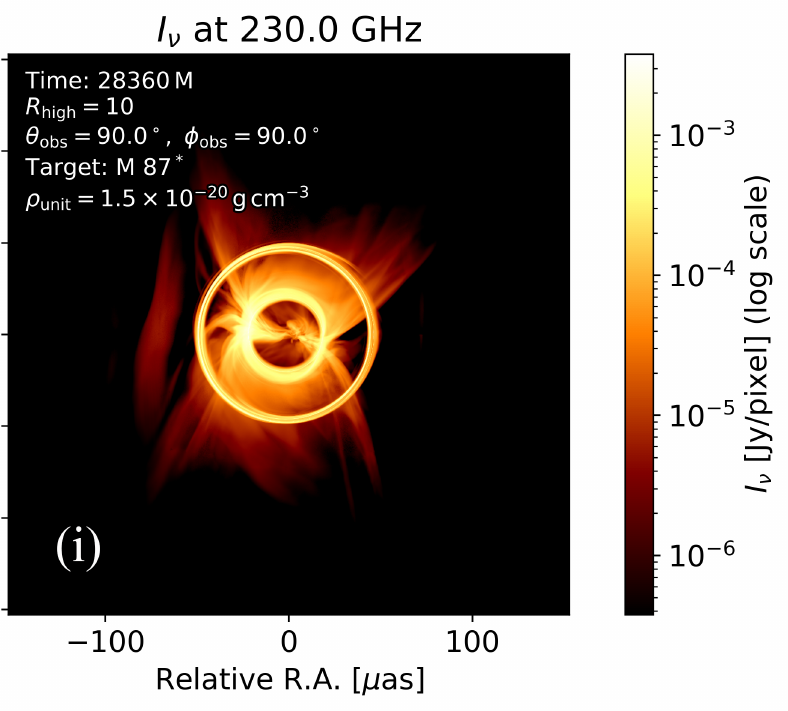}
\caption{Synthetic horizon-scale total-intensity ($I_\nu$) images at $230\,\mathrm{GHz}$.
Panels (a--c): run~$\tt{VT}$ with $\phi_{\rm obs}=90^\circ$ at three representative times.
Movie: \href{https://youtu.be/Z201nzWcnss}{https://youtu.be/Z201nzWcnss}.
Panels (d--f): run~$\tt{VT}$ with $\phi_{\rm obs}=0^\circ$ at the corresponding epochs.
Movie: \href{https://youtu.be/R_2J4vxRqgg}{https://youtu.be/R\_2J4vxRqgg}.
Panels (g--i): run~$\tt{CP}$ with $\phi_{\rm obs}=90^\circ$ at three representative times.
Movie: \href{https://youtu.be/1NRrb8LMbWI}{https://youtu.be/1NRrb8LMbWI}.
The color scale shows $I_\nu$ in $\rm Jy\,pixel^{-1}$ (logarithmic), and each panel is labeled by its snapshot time and viewing geometry.}
\label{fig:BH_shadow}
\end{figure*}

In Fig.~\ref{fig:BH_shadow}, selected synthetic images of the BH shadows at different phase angles are presented. Panels (a)-(c) illustrate the process as the secondary BH approaches and enters the torus. Due to the large mass ratio ($q = 0.1$) of the two BHs, the secondary BH shadow is much smaller than the primary one but still visible as a bright spot. As it approaches and enters the torus, the interaction leads to the formation of a luminous bow shock, which then evolves into a trailing tail. This luminous feature is produced by compression and heating in the secondary-driven shock. The related plasma diagnostics are discussed in Section~\ref{sec:shock_3D}. 
Although the bow shock is clearly visible in GRMHD simulations, the contribution to total emission from the secondary BH is much weaker than that of the primary BH at submillimeter frequencies (see Fig.~\ref{fig:lc_BH2}a and c). 

In the middle row (panels (d)-(f)), we show the same set of snapshots but for a different LOS, with $\phi_{\rm obs}=0^\circ$. In this viewing angle, the observer becomes nearly aligned with the two BHs during the orbit, and strong gravitational lensing can occur. In particular, at $t=27{,}840\,\rm M$ the alignment is favorable for significant magnification of compact emission near the BHs, producing a pronounced transient brightening in the synthetic image. This lensing event corresponds directly to the luminous flare seen in the NIR light curve in Fig.~\ref{fig:lc_BH2}b.

Finally, the lower middle row (panels (g)-(i)) of Fig.~\ref{fig:BH_shadow} presents synthetic images from the run~$\tt{CP}$. In this coplanar configuration, the secondary BH remains embedded within the torus and periodically passes from the foreground to the background relative to the primary along the LOS (panel~(h) shows the secondary in front, while panel~(i) shows it behind and correspondingly brighter). These passages create favorable alignment geometries that produce strong gravitational lensing, leading to pronounced, time-dependent distortions and magnification of the emission from the inner flow.

The corresponding orbit-folded light curves of run~$\tt{CP}$ (Fig.~\ref{fig:lc_BH2}e and f) reveal a distinct difference between the sub-millimeter and NIR bands.
At 230~GHz, the total flux is dominated by the accretion flow of the primary BH. 
However, when the secondary BH passes between the primary BH and the observer (orbital phase $\simeq 0.2$-$0.3$), the gravitational lensing effect becomes extreme. The secondary BH strongly magnifies the primary BH's photon ring emission, producing a pronounced brightening and a characteristic double-peaked profile (see Fig.~\ref{fig:lc_BH2}e). This scenario corresponds to the shadow image in Fig.~\ref{fig:BH_shadow}h, which displays a highly twisted photon ring and lensed image of the primary BH.
In contrast, at 138~THz (Fig.~\ref{fig:lc_BH2}f), the emission hierarchy is reversed. The secondary BH becomes the dominant source of radiation due to the significantly higher electron temperatures in its vicinity, while the emission coming from the accretion flow onto the primary BH fades. Consequently, the peak flux occurs when the secondary BH passes behind the primary BH (orbital phase $\sim 0.72$). In this configuration, the more massive primary BH lenses the compact emission of the secondary BH, focusing it into a prominent Einstein ring (as seen in Fig.~\ref{fig:BH_shadow}i) and generating a sharp localized flare. 

Figure~\ref{fig:BH_shadow}g also shows a mini-jet launched from the secondary BH, tracing gravitationally unbound, magnetized plasma.
The twisted, helical magnetic field lines around the secondary are discussed in Section~\ref{sec:shock_3D}.
A related observational motivation comes from OJ~287, where secondary-jet signatures have been discussed through jet-position-angle modeling and high-resolution RadioAstron imaging \citep{2021MNRAS.503.4400D,2025ApJ...992..110V}.
As the secondary BH orbits, the outflow and jet are sheared and twisted, and can feed back on the surrounding gas \citep{2015PhRvD..91h4044P,2019PhRvD.100f3009C}. Hydrodynamic studies show that sufficiently strong outflows can disrupt the gravitational wake and reduce, or even reverse, the effective drag force (``negative dynamical friction'') \citep{2020MNRAS.494.2327L}, and that interacting outflows in binaries can yield a net positive torque that promotes orbital expansion \citep{2022ApJ...932..108W}. By analogy, an orbit-modulated outflow from the secondary BH may alter disk torques and partially offset inward migration, which we leave to investigate in future work.

\subsection{Dynamical Evolution of BH Spin and Disk Orientation}
\label{sec:orientation}

\begin{figure*}[!t]
    \centering
    \includegraphics[width=0.45\linewidth]{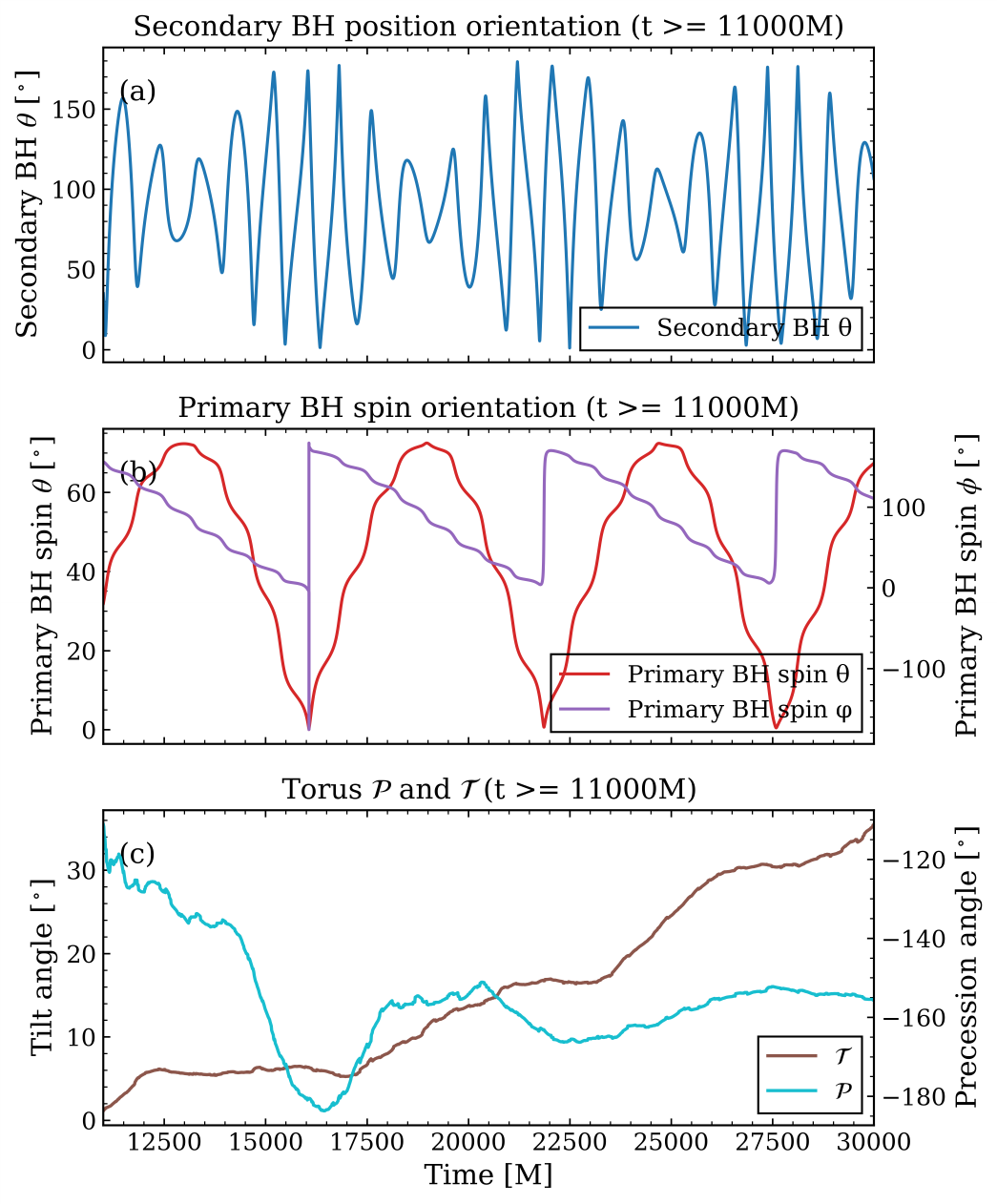}
    \includegraphics[width=0.53\linewidth]{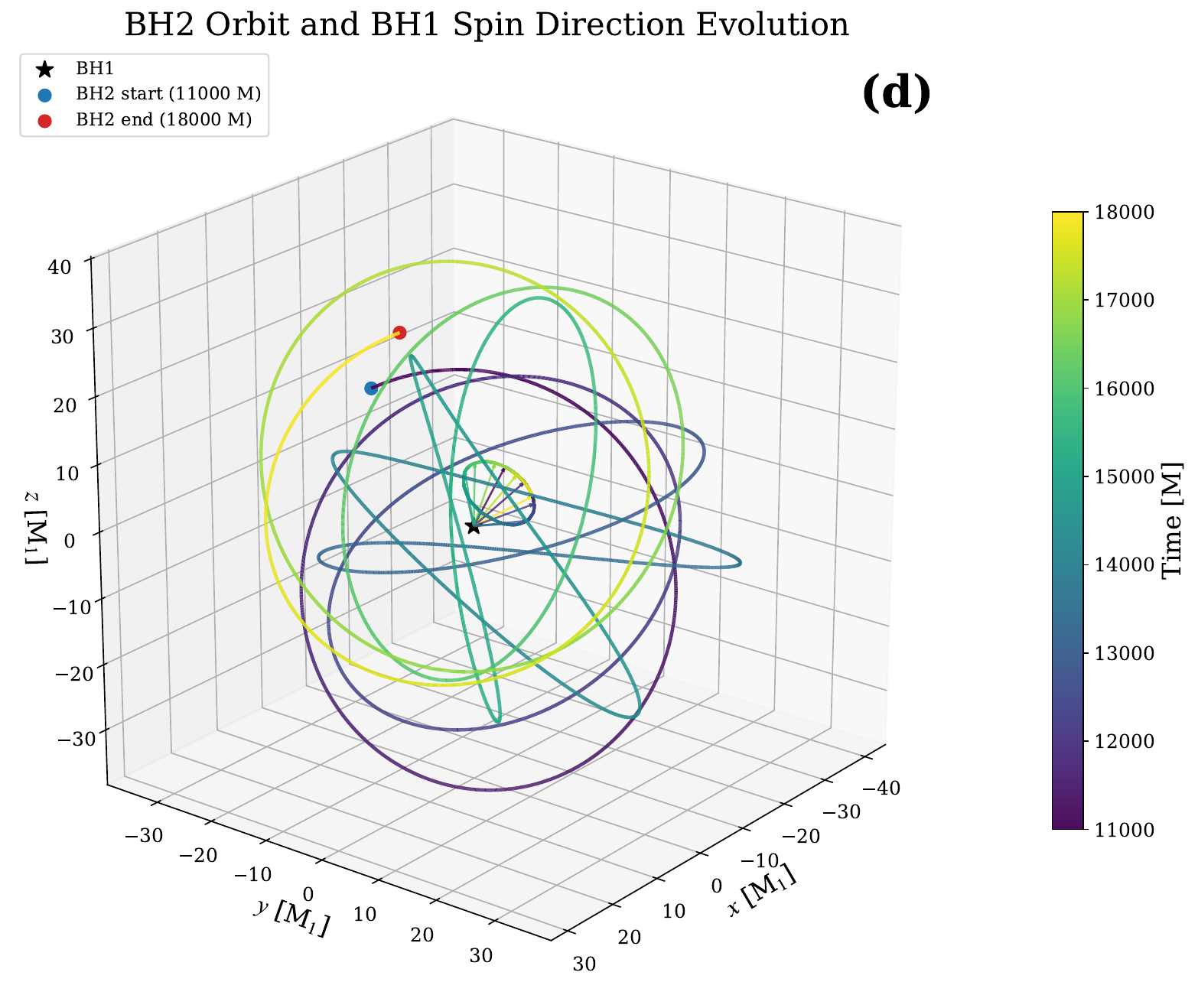}
    \caption{Orientation evolution in run~$\tt{EP}$ for $t\ge 11{,}000\,\rm M$.
(a) Polar angle $\theta$ of the secondary BH position vector.
(b) Primary BH spin orientation: polar angle $\theta_{\rm spin}$ (red; left axis) and azimuthal angle $\phi_{\rm spin}$ (purple; right axis).
(c) Torus angular-momentum orientation: tilt $\mathcal{T}$ (brown; left axis) and precession angle $\mathcal{P}$ (cyan; right axis), measured following \citet{2005ApJ...623..347F, 2025ApJ...995..112J}.
(d) 3D evolution of the secondary BH orbit together with the primary BH spin direction: the colored curve traces the secondary BH trajectory, the short vector anchored at the primary BH indicates the instantaneous primary BH spin direction.}
    \label{fig:orientation}
\end{figure*}

To validate the physical origin of the precessing jet morphology observed in run~$\tt{EP}$, we quantify the time evolution of the position and orientations of the BHs and torus.
Fig.~\ref{fig:orientation} presents the position angles for the secondary BH orbit, the primary BH spin, and the torus angular momentum.

The first panel of Fig.~\ref{fig:orientation} tracks the polar angle $\theta$ of the secondary BH's position vector. The secondary BH is initialized on a highly inclined orbit with $i=90^{\circ}$ relative to the initial torus midplane. As the system evolves ($t \ge 11{,}000\,\rm M$), $\theta$ develops large-amplitude oscillations, indicating strong nodal precession of the orbital plane driven by LT torques from the spinning primary BH. As a result, the orbit cyclically reorients, alternating between phases that are closer to vertical disk crossings and phases that are more nearly coplanar with the disk.

The second panel shows the evolution of the primary BH's spin orientation, characterized by the polar angle $\theta_{\rm spin}$ and azimuthal angle $\phi_{\rm spin}$. Although the primary BH is much more massive than the disk and the secondary BH, the cumulative torque from the orbiting secondary BH induces a secular drift in the spin axis over long timescales. The spin axis precesses, with $\theta_{\rm spin}$ and $\phi_{\rm spin}$ showing coupled oscillations. 

Panel c shows the time evolution of the torus angular-momentum orientation, characterized by the tilt angle $\mathcal{T}$ and precession angle $\mathcal{P}$, measured using the same procedure as in \citet{2025ApJ...995..112J}. The tilt of the torus increases secularly from a few degrees at $t\simeq11{,}000\,\rm M$ to $\mathcal{T}\sim30^\circ$ by $t\simeq30{,}000\,\rm M$, indicating a gradual growth of the global misalignment relative to the $+z$ axis. Meanwhile, $\mathcal{P}$ exhibits a pronounced early-time sweep followed by slower late-time evolution, suggesting that azimuthal reorientation becomes less rapid after the initial adjustment phase. In comparison of panels (b) and (c), the torus orientation tends to evolve in the same sense as the primary BH spin reorientation, consistent with the torus responding to the changing torque geometry imposed by the primary BH. However, the characteristic time scale of the primary BH spin precession is noticeably shorter: $\theta_{\rm spin}$ and $\phi_{\rm spin}$ vary rapidly and coherently, whereas $\mathcal{T}$ and $\mathcal{P}$ change more gradually, implying that the torus responds more slowly and cannot track the fastest components of the primary BH spin precession.

Panel (d) visualizes the coupled 3D evolution by overlaying the secondary BH orbital trajectory with the instantaneous primary BH spin direction. From $t\simeq11{,}000$--$18{,}000\,\rm M$, the secondary BH follows a highly variable path, oscillating between the vertical and coplanar phases (consistent with Fig.~\ref{fig:orientation}a). 
During this interval, the primary BH spin precesses around a tilted axis, increasing the torus tilt angle seen in panel (c). This indicates that the torus angular momentum is gradually torqued to follow the evolving BH configuration, although its longer response time ensures that its precession lags behind the primary BH spin precession.

\subsection{Morphology and Observational Signatures of the Precessing Binary}
\label{sec:run4_results}
Similar to \citet{2024ApJ...967...70R}, spin-orbit coupling in our run~$\tt{EP}$ drives a time-dependent reorientation of the primary BH spin axis, which in turn warps and twists the inner disk and jet orientation. To capture a more realistic and strongly time-dependent binary environment, run~$\tt{EP}$ adopts a high-spin configuration for both BHs and an eccentric, highly inclined orbit. 

\begin{figure*}[!t]
\centering 	
\includegraphics[height=.33\linewidth]{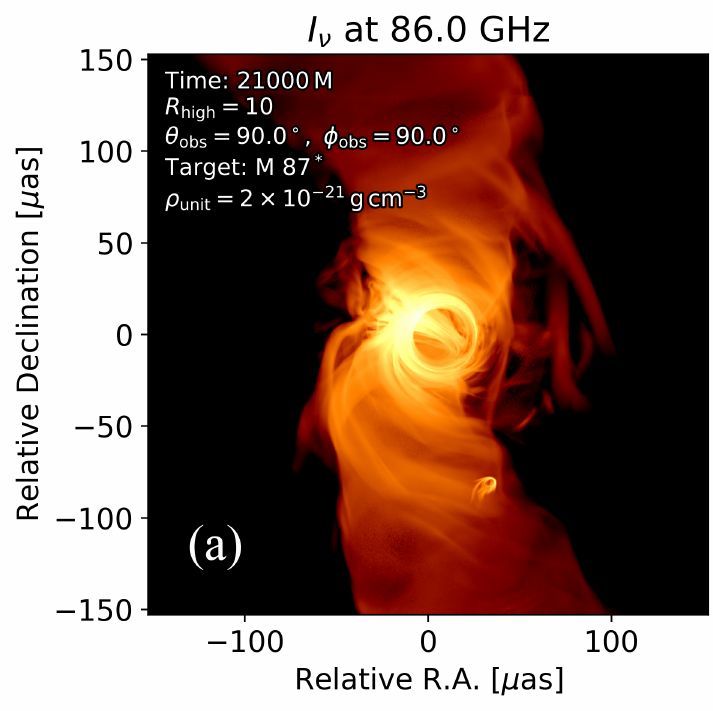}
\includegraphics[height=.33\linewidth]{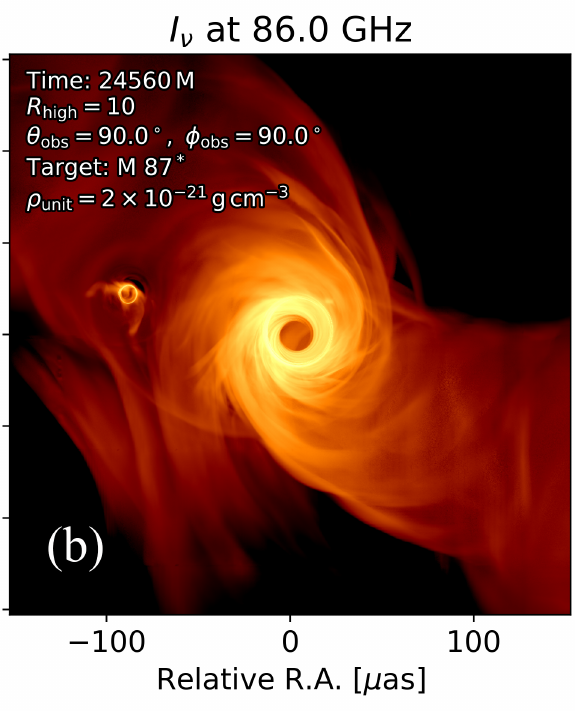}
\includegraphics[height=.33\linewidth]{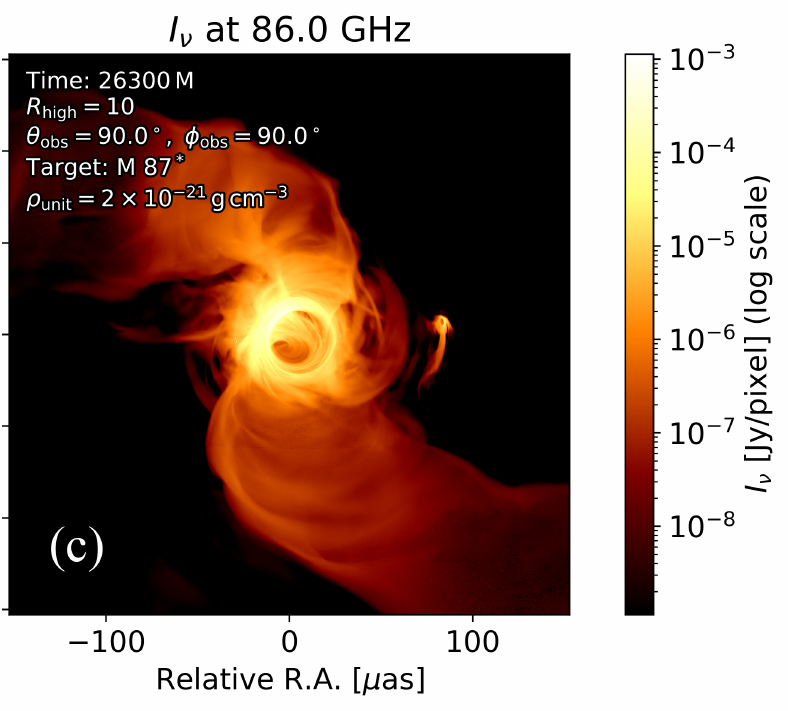}
\caption{Synthetic total-intensity ($I_\nu$) images from run~$\tt{EP}$ at $86\,\mathrm{GHz}$ at three representative times. Panel (a) $t=21{,}000\,\rm M$; Panel (b) $t=24{,}560\,\rm M$; Panel (c) $t=26{,}300\,\rm M$.
The color scale shows $I_\nu$ in $\mathrm{Jy\,pixel^{-1}}$ (logarithmic), and each panel is annotated with the snapshot time and viewing geometry. Movie: \href{https://youtu.be/dOQ8ppdLvWU}{https://youtu.be/dOQ8ppdLvWU}.}
\label{fig:precessing_jet}
\end{figure*}

Unlike runs~$\tt{VT}$ and $\tt{CP}$, the spinning primary BH in run~$\tt{EP}$ induces strong frame dragging, and the associated LT torque drives pronounced precession of the secondary BH orbit. As the secondary BH circulates, spin-orbit coupling also causes the primary BH spin axis to vary in time (see Section~\ref{sec:orientation} for a quantitative analysis of the disk and spin orientation evolution), leading to twisted and precessing jet structure.
Fig.~\ref{fig:precessing_jet} shows synthetic $86\,{\rm GHz}$ images at three representative times, showing a changing jet orientation. The jet is bent and twisted, with a visibly wobbling nozzle that produces a time-dependent asymmetric emission pattern. Such precessing jet morphologies provide a generic diagnostic of spin--orbit coupling in MAD-like binaries, although applying this mechanism directly to luminous blazars such as OJ~287 would require geometrically thin-disk/radiative physics beyond the present models \citep[e.g.,][]{2022ApJ...932...72Z, 2026A&A...705A..23G}. 

\begin{figure}[!t]
\centering 	
\includegraphics[width=\linewidth]{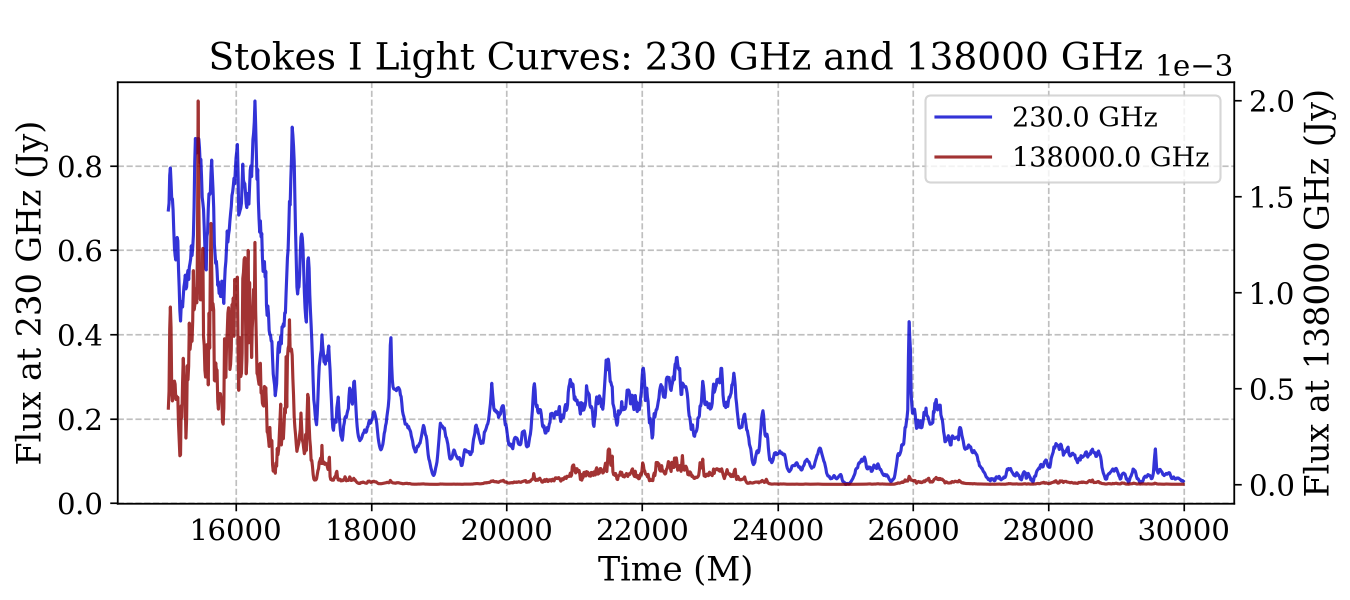}
\caption{Run~$\tt{EP}$ Stokes-$I$ light curves at $230\,\mathrm{GHz}$ (blue; left axis) and $1.38\times10^{5}\,\mathrm{GHz}$ (brown; right axis).
}
\label{fig:lc_run3}
\end{figure}

Fig.~\ref{fig:lc_run3} shows the light curves of the run~$\tt{EP}$ at $230\,{\rm GHz}$ and $138\,{\rm THz}$. Despite the pronounced jet and inner-flow reorientation, the thermal synchrotron emission does not show the regular quasi-periodic flaring seen in controlled runs~$\tt{VT}$ and $\tt{CP}$. Instead, the variability is largely stochastic. A bright self-lensing flare appears only near $t\approx26{,}000\,\rm M$, when the secondary BH passes behind the primary BH and becomes nearly aligned with the LOS. At other epochs, shock activity can enhance emission, as in run~$\tt{VT}$, but these events are typically too weak to generate clearly separated spikes in the light curves.

\section{Summary and Discussion}
In this work, we perform global 3D GRMHD simulations coupled with GRRT calculations to investigate the dynamics of accretion flows and the observable signatures of SMBBHs embedded in MAD flows. Using an approximate time-dependent binary spacetime, we capture horizon-scale physics for a moderate mass-ratio system ($q=0.1$). Our main findings are as follows.
\begin{enumerate}
    \item \textit{Dynamics--observable decoupling:} Even when the secondary BH produces strong, periodic shock activity and accretion-rate bursts in the vertical-impact run~$\tt{VT}$, the thermal synchrotron emission does not always exhibit corresponding bright flares. The light curves are instead often dominated by stochastic MAD variability, with clear periodic features emerging mainly during favorable self-lensing alignments or rare episodes of strong shock heating.
    \item \textit{Frequency-dependent emission hierarchy:} The relative importance of the two BHs depends strongly on observing frequencies. At sub-millimeter wavelengths, the primary MAD flow dominates the total flux, while at NIR and higher frequencies, the hotter, more compact plasma near the secondary BH can become much more visible.
    \item \textit{Self-lensing flares:} Coplanar binaries (run~$\tt{CP}$) produce frequent, narrow, high-contrast flares when the two BHs approach LOS alignment. At $230$~GHz, the secondary BH can lens primary-flow emission, whereas in the NIR, the primary BH can strongly magnify compact emission associated with the secondary one.
    \item \textit{Shock-injected nonthermal emission:} A phenomenological shock-injected power-law electron tail mainly affects the high-frequency SED. In run~$\tt{VT}$, this enhancement is associated with the high-$\chi_{\rm shock}$ compression around the secondary BH, while run~$\tt{CP}$ lacks a comparable shock volume and its sharp flares remain primarily lensing-driven.
    \item \textit{Jet precession:} In the high-spin, eccentric run~$\tt{EP}$, torques from the secondary BH drive time-dependent reorientation of the primary BH spin, producing a wobbling, twisted jet morphology that may be a useful generic signature of spin--orbit coupling in thick-disk SMBBHs, such as LLAGN.
\end{enumerate}

Our GRMHD results are broadly consistent with the recent studies of \citet{2024ApJ...967...70R} and \citet{2025ApJ...979L..24R}: in all cases, the secondary BH perturbs the MAD flow through localized shocks, enhanced secondary accretion, magnetized outflows, and orbit-modulated variability, while the primary MAD turbulence can obscure clean periodic signatures. The main distinction of the present work is the observable post-processing: we ray-trace the simulations on the same time-dependent binary metric to separate shock-powered thermal emission from self-lensing and to quantify the frequency-dependent transition between primary-dominated sub-millimeter emission and secondary-dominated NIR emission. We also distinguish our geometrically thick, radiatively inefficient MAD models from the OJ~287-motivated geometrically thin-disk calculations of \citet{2025ApJ...993L..22R}, which include radiative cooling physics; those simulations are more directly applicable to luminous thin-disk impacts, whereas the present calculations target LLAGN-like, geometrically thick MAD systems.

Taken together, these results suggest that identifying SMBBHs will require combining time-domain diagnostics, such as short self-lensing spikes, with high-angular-resolution imaging of the inner accretion flows and jets. Next-generation VLBI facilities such as the EHT/ngEHT \citep{2023Galax..11...61J}, together with coordinated multi-wavelength monitoring, will be important for separating binary signatures from intrinsic variability in a single active galactic nucleus \citep{2025LRR....28....4A}, while the next-generation Very Large Array (ngVLA) will provide complementary sensitivity to larger-scale jet structure and possible signatures of jet precession.

However, for distant candidates of SMBBHs, directly resolving the two components may remain challenging. For illustrative geometrically thin-disk/blazar candidates such as OJ~287, expected angular separations can be only a few microarcseconds, below the $\sim 10$--$20\,\mu\mathrm{as}$ resolution of current and near-term ground-based mm-VLBI. In this under-resolved regime, time-domain diagnostics may be more powerful: binary self-lensing can generate narrow, approximately symmetric flares near conjunction, but becomes prominent mainly for nearly edge-on viewing. Because luminous blazars are jet-dominated and may be governed by different disk thermodynamics than our radiatively inefficient MAD models, this comparison should be regarded only as an observational motivation rather than a direct application. Extending baselines beyond Earth via future space-VLBI concepts, such as the Black Hole Explorer \citep{2024SPIE13092E..2DJ}, could ultimately connect timing-based evidence to direct spatial constraints.

We note several limitations. Our main GRRT analysis still uses thermal synchrotron emission, while the nonthermal calculation in Section~\ref{sec:nonthermal_effects} is a phenomenological post-processing test rather than a self-consistent particle-acceleration model. We also omit inverse Compton scattering, which can contribute to shock-powered high-energy emission. We also focus on geometrically thick, radiatively inefficient flows. Thin-disk systems such as OJ~287 \citep{2019ApJ...882...88V} may exhibit different shock and variability properties. Finally, we selected edge-on configurations to maximize lensing: a broader inclination survey is needed to quantify detectability across the wider SMBBH population. 

\appendix

\section{Validation Tests}
\label{app:validation}

\subsection{Single-BH GRMHD Evolution and Cartesian Kerr--Schild Metric Limit}
\label{app:single_bh_validation}

\begin{figure*}[!t]
\centering
\includegraphics[width=0.96\linewidth]{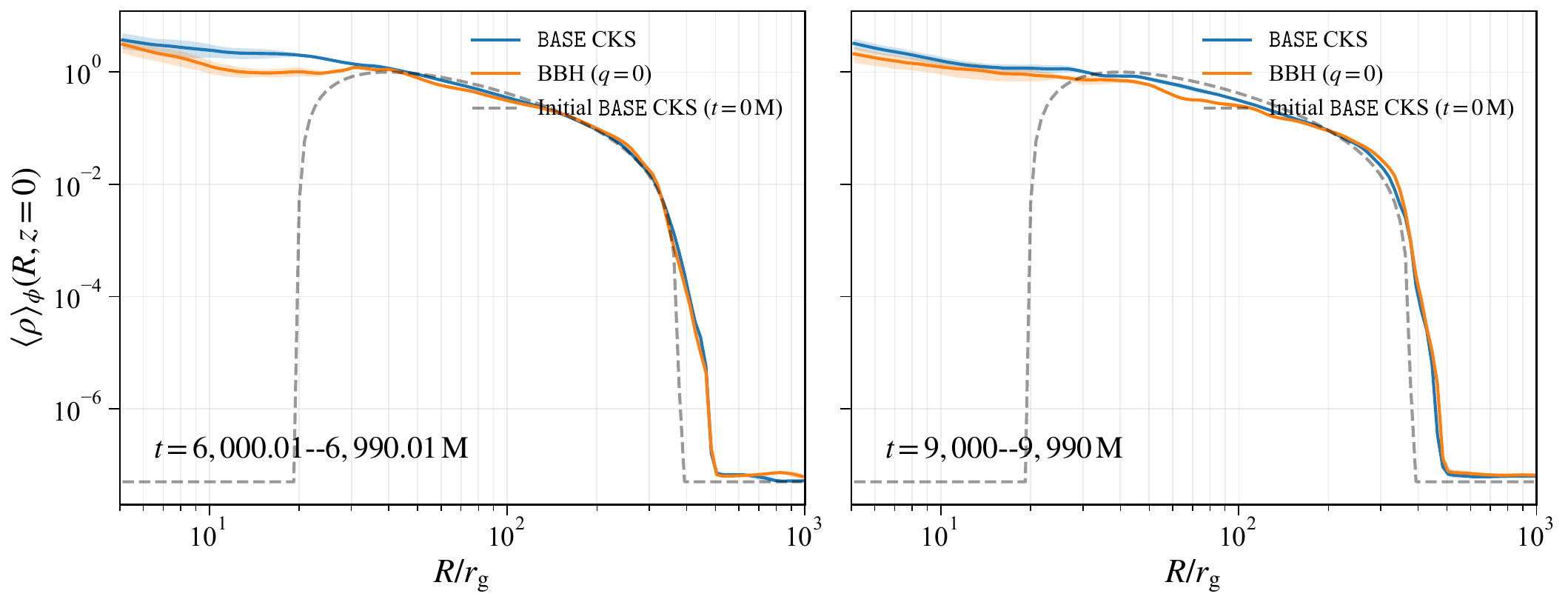}
\caption{Azimuthally-averaged density profiles at midplane for the single-BH validation. Solid curves show the time-averaged profiles from the standard single-BH Cartesian Kerr--Schild run (blue) and from the BBH calculation with $q=0$ (orange); shaded bands show the one-standard-deviation temporal scatter over each averaging interval. The black dashed curve shows the initial Cartesian Kerr--Schild profile at $t=0\,\rm M$. Left: $t=6{,}000.01$--$6{,}990.01\,\rm M$. Right: $t=9{,}000$--$9{,}990\,\rm M$.}
\label{fig:single_bh_validation}
\end{figure*}

We test the single-BH limit of the approximate BBH metric. Setting $q=0$ makes the secondary mass vanish, thus, the metric should be reduced to the Cartesian Kerr--Schild (CKS) form for an isolated Schwarzschild BH. We verify this algebraically by sampling $20{,}000$ points outside the horizon. The maximum component-wise difference between the binary metric with $q=0$ and the analytic CKS metric is $1.1\times10^{-15}$, and the inverse-metric residual is $\max |g_{\mu\nu}g^{\nu\lambda}-\delta_\mu^\lambda|=6.0\times10^{-16}$. These values show that the binary metric reduces to the single-BH CKS metric to roundoff accuracy when the secondary mass is set to zero.

We also compare the corresponding GRMHD evolution between the standard single-BH CKS run and from the BBH calculation with $q = 0$ seen in Fig.~\ref{fig:single_bh_validation}. The standard CKS run and the $q=0$ BBH run produce the same broad midplane density structure, including the torus body and its outer radial decline, over both averaging intervals. Small differences in the profiles are expected because the two calculations follow independent evolution paths and use different GRMHD infrastructure: the $q=0$ BBH run uses the \texttt{DynGRMHD} module, whereas the standard CKS run uses the single-spacetime \texttt{GRMHD} module \citep{Stone2024,2025ApJS..278...50Z}. The agreement of the azimuthally averaged density profiles nevertheless confirms that the $q=0$ BBH run evolves the same single-BH torus solution rather than introducing a spurious binary-metric effect.

\subsection{GRRT Metric and Flux Consistency}
\label{app:grrt_metric_validation}

\begin{figure*}[!t]
\centering
\includegraphics[width=0.96\linewidth]{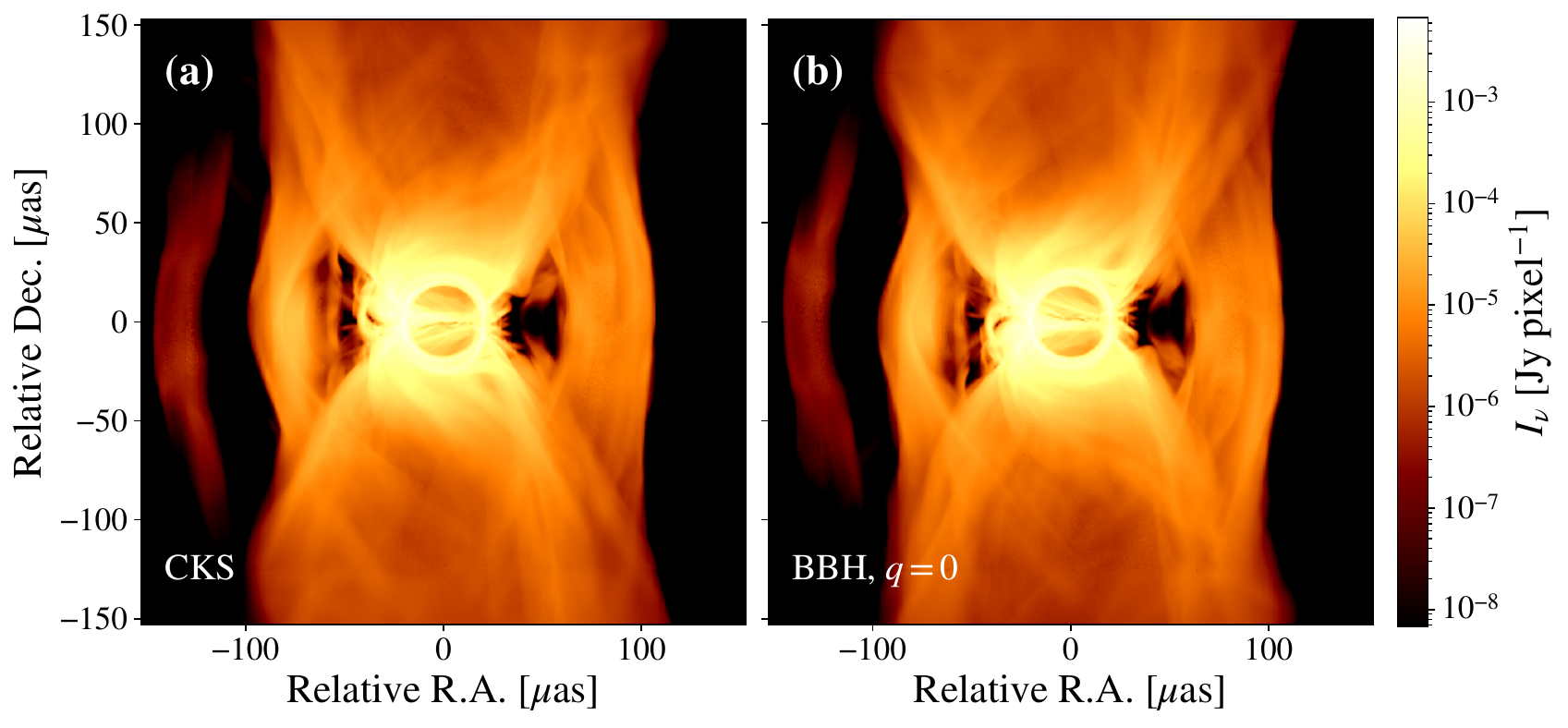}
\caption{230~GHz GRRT consistency test for the single-BH limit. Panels (a) and (b) show the same single-BH torus snapshot ray-traced images using the analytic CKS metric (left) and the $q=0$ BBH metric (right), respectively. The integrated flux comparison is discussed in the text.}
\label{fig:grrt_metric_validation}
\end{figure*}

We also validate the GRRT implementation in the single-BH limit by ray-tracing the same GRMHD snapshot with two independent metric paths. In Fig.~\ref{fig:grrt_metric_validation}a, photons are propagated on the analytic CKS spacetime. In panel b, the same snapshot is ray-traced using the BBH metric with $q=0$. As shown in Fig.~\ref{fig:grrt_metric_validation}, the resulting images are visually consistent. At $230\,{\rm GHz}$, the total Stokes-$I$ fluxes are $6.469\,{\rm Jy}$ and $6.485\,{\rm Jy}$ with an absolute flux difference of $1.60\times10^{-2}\,{\rm Jy}$. 
Although both calculations represent the same continuum spacetime in the $q=0$ limit, perfect pixel-to-pixel agreement is not expected. Because the two calculations use independent ray-initialization and metric-evaluation paths. The negligible flux discrepancy demonstrates that GRRT observables are insensitive to the choice between standard CKS and $q=0$ BBH metric implementations.

\subsection{Shock-Injected Nonthermal GRRT comparison}
\label{app:nonthermal_image_validation}

\begin{figure*}[!t]
\centering
\includegraphics[width=0.96\linewidth]{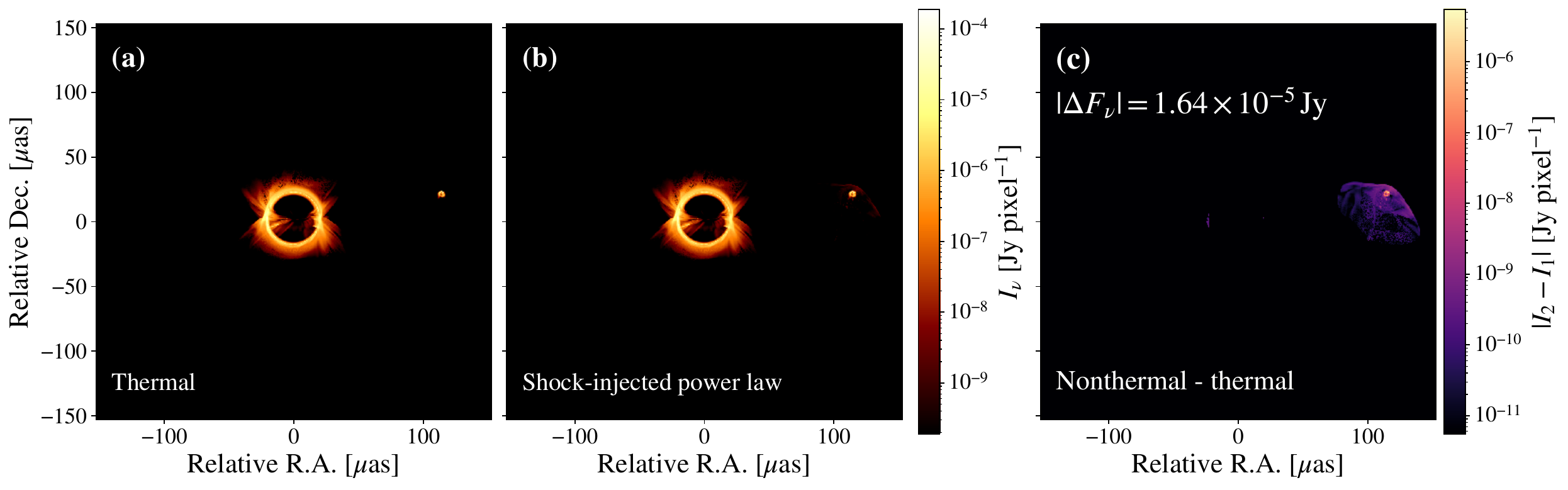}
\caption{Comparison of the thermal-only synthetic image and the shock-injected nonthermal GRRT image for run~$\tt{VT}$ at $138\,{\rm THz}$. Panel (a): thermal-only GRRT post-processing at $t=16{,}280\,\rm M$. Panel (b): same snapshot with shock-injected nonthermal electrons included. Panel (c) shows the absolute image difference. The nonthermal prescription reveals localized emission from the secondary-driven shock front, but the integrated flux increase is only $1.64\times10^{-5}\,{\rm Jy}$.}
\label{fig:nonthermal_image_validation}
\end{figure*}

\begin{table*}[!t]
\centering
\caption{GRRT image-resolution check for the same snapshot shown in Fig.~\ref{fig:nonthermal_image_validation}. The table lists the total Stokes-$I$ flux at each image resolution and the percent resolution-to-resolution difference.}
\label{tab:grrt_image_resolution}
\begin{tabular}{ccccc}
\hline \hline
Model & $\nu$ (GHz) & $F_{\nu,1000^2}$ (Jy) & $F_{\nu,2000^2}$ (Jy) & $100|F_{\nu,2000^2}-F_{\nu,1000^2}|/F_{\nu,1000^2}$ \\
\hline
Thermal & $86$ & $4.5827$ & $4.5823$ & $0.0091\%$ \\
Thermal & $230$ & $8.6853$ & $8.6820$ & $0.0376\%$ \\
Thermal & $138000$ & $4.8316\times10^{-3}$ & $5.0216\times10^{-3}$ & $3.93\%$ \\
Nonthermal & $86$ & $4.5824$ & $4.5820$ & $0.0085\%$ \\
Nonthermal & $230$ & $8.6853$ & $8.6818$ & $0.0407\%$ \\
Nonthermal & $138000$ & $4.8480\times10^{-3}$ & $5.0379\times10^{-3}$ & $3.92\%$ \\
\hline
\end{tabular}
\end{table*}

To quantify the impact of shock-injected nonthermal electrons on the GRRT images, we examine a representative snapshot of run~$\tt{VT}$ at $t=16{,}280\,\rm M$. Figure~\ref{fig:nonthermal_image_validation} compares the thermal-only and nonthermal GRRT synthetic images for this snapshot. The power-law component is related to the local shock diagnostic $\chi_{\rm shock}$, so the greatest change appears in the compact shocked structure associated with the secondary BH. The difference image makes this shock-front emission visible, demonstrating that the post-processing responds to the intended compressive region.

The global contribution of this component is nevertheless small in this snapshot. At $138\,{\rm THz}$, the total Stokes-$I$ flux changes from $4.8316\times10^{-3}\,{\rm Jy}$ in the thermal calculation to $4.8480\times10^{-3}\,{\rm Jy}$ when the shock-injected power-law tail is included, a very small increase of $1.64\times10^{-5}\,{\rm Jy}$. We repeated this comparison with a $2000^2$ image grid at $86$, $230$, and $138000\,{\rm GHz}$. The resulting total fluxes and percent resolution-to-resolution differences are listed in Table~\ref{tab:grrt_image_resolution}. The total fluxes in the millimeter-band change by $\lesssim0.05\%$ when the image resolution is doubled, while the $138\,{\rm THz}$ total flux changes by $\simeq4\%$. Thus, the nonthermal prescription can reveal localized shock-front emission and modestly extend the image dynamic range. However, as also shown in Fig.~\ref{fig:nonthermal_sed_phase}, the largest contribution occurs at higher frequencies, e.g. $10^{15}$--$10^{18}\,\rm Hz$.

\subsection{Resolution Sensitivity of the Secondary-Local Outflow}
\label{app:secondary_resolution}

\begin{figure*}[!t]
\centering
\includegraphics[width=0.96\linewidth]{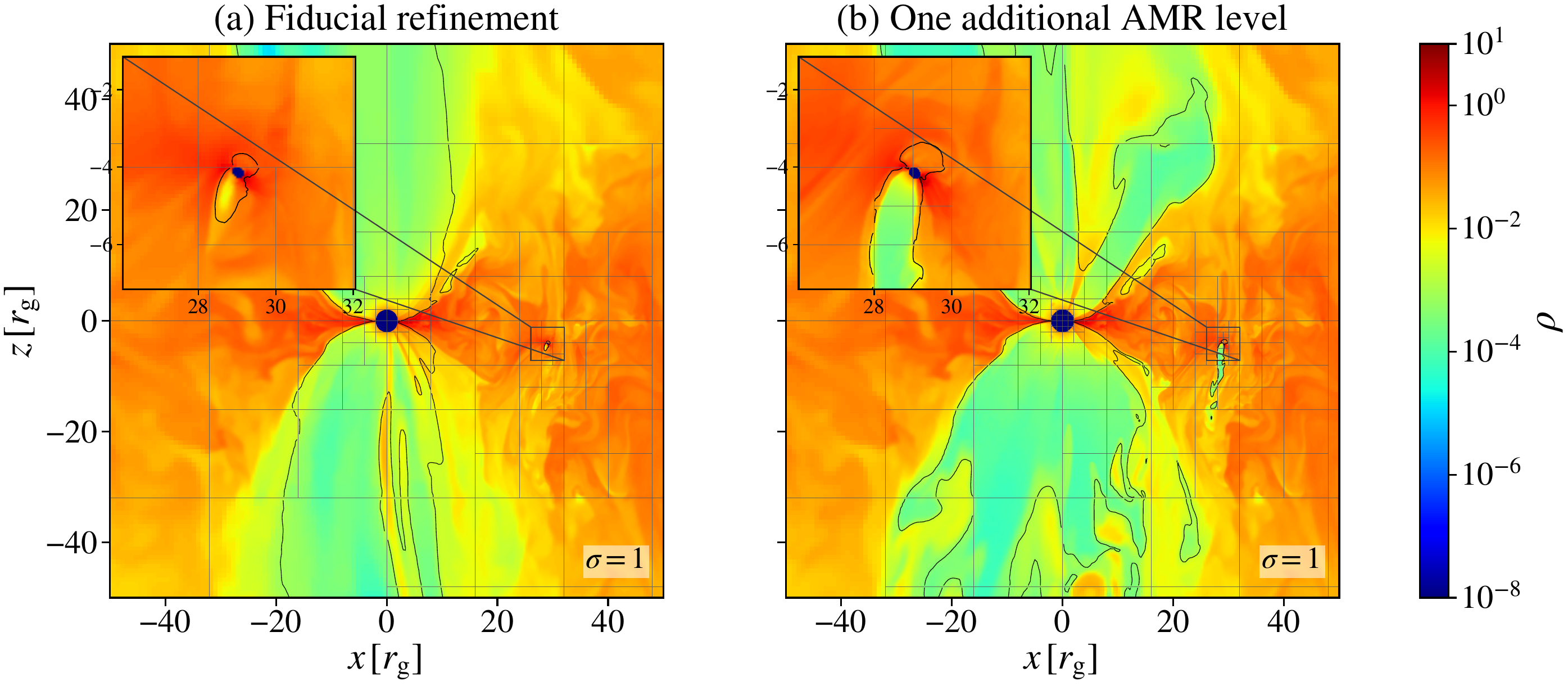}
\caption{Resolution comparison near the secondary BH in run~$\tt{VT}$, shown in an $x$--$z$ slice at $t=26{,}590\,\rm M$. Panel (a) shows the fiducial production resolution, while panel (b) shows the same calculation restarted from the fiducial run~$\tt{VT}$ snapshot at $t=26{,}400\,\rm M$ with one additional AMR level around the secondary. The color scale gives density, the black contours mark $\sigma=1$, and the insets zoom into a $\pm3\,r_{\rm g}$ region centered on the secondary BH. The additional refinement makes the low-density, magnetized channel near the secondary clearer and more continuous.}
\label{fig:secondary_resolution_comparison}
\end{figure*}

\begin{figure}[!t]
\centering
\includegraphics[width=.7\linewidth]{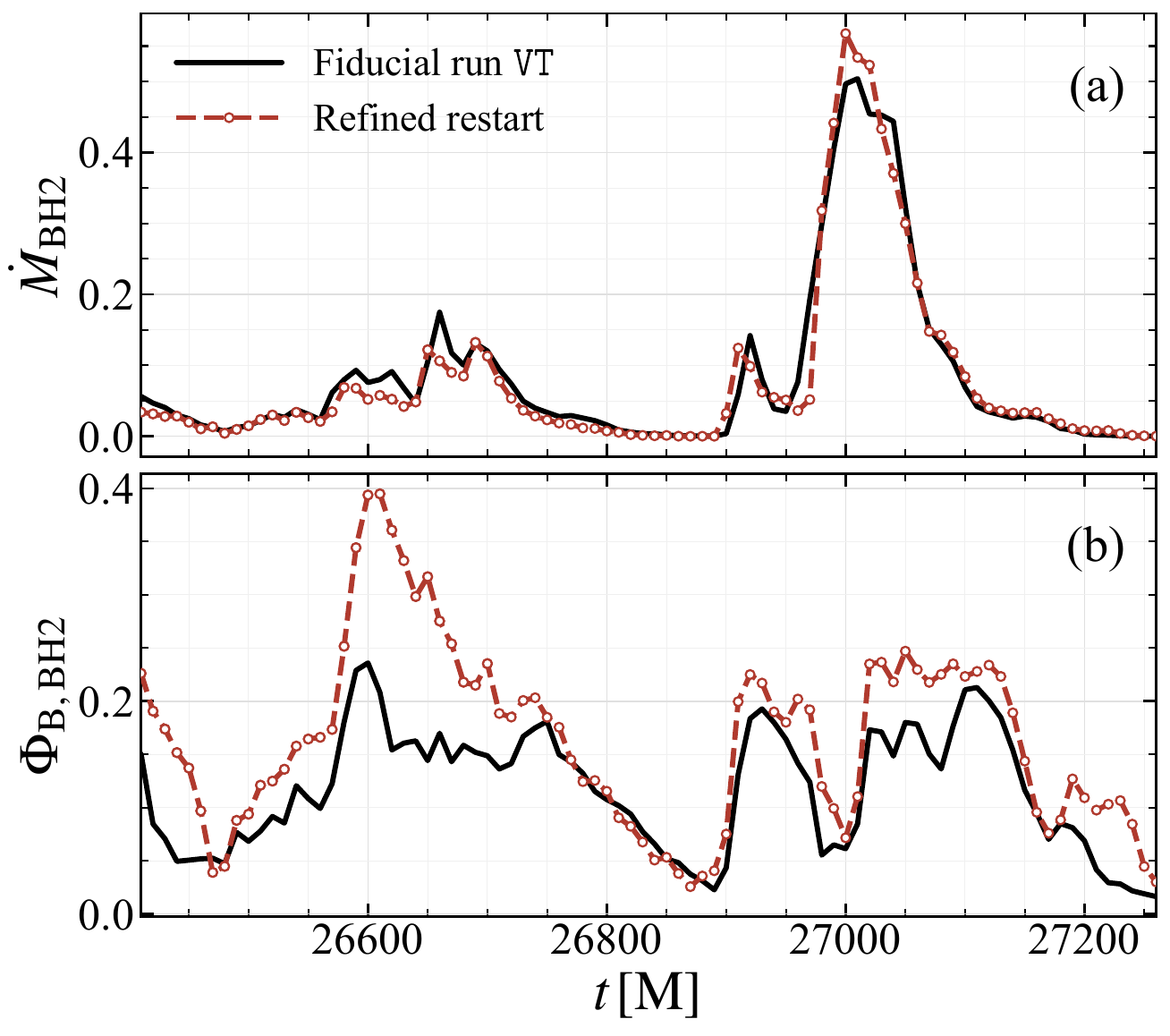}
\caption{Mass accretion rate (upper) and magnetic flux rate (lower) at the horizon of secondary-BH in the fiducial run~$\tt{VT}$ and in the refined restart over the common interval $t=26{,}410$--$26{,}700\,\rm M$. }
\label{fig:bh2_resolution_diagnostics}
\end{figure}

Figure~\ref{fig:secondary_resolution_comparison} presents a resolution test of the outflow/jet morphology from the secondary BH in the run~$\tt{VT}$ after restarting from $t=26{,}400\,\rm M$ with one additional AMR level. In the fiducial-resolution run~$\tt{VT}$, the spinning secondary BH horizon is resolved by $N_{H,2}\simeq5.4$ finest cells across its diameter; the refined restart doubles it to $N_{H,2}\simeq10.8$. The large-scale torus and environment surrounding the secondary BH are qualitatively similar between the two snapshots, although modest differences arise from the nonlinear nature of MHD turbulence. The higher-resolution calculation reveals a more distinct low-density jet-like region near the secondary and produces a sharper $\sigma=1$ contour. For comparison, \citet{2024ApJ...967...70R} used $N_{H,2}\sim4$ for their nonspinning secondary BH with $q=0.1$. Our fiducial run therefore has about $35\%$ more cells across the horizon diameter of the secondary BH, while the refined restart has about $2.7$ times more. Nevertheless, Fig.~\ref{fig:secondary_resolution_comparison} shows that even this fiducial resolution is not sufficient to claim convergence of the secondary-local jet morphology or power. The refined restart, therefore, provides a useful check that additional resolution makes the magnetized low-density channel clearer. However, our main conclusions rely on the global disk response, shock morphology, primary-dominated MAD variability, and self-lensing geometry rather than on converged secondary jet power.

The refined restart preserves the timing and approximate amplitude of $\dot{M}_{\rm BH2}$ (Fig.~\ref{fig:bh2_resolution_diagnostics}a), with mean and peak values $0.80$ and $0.76$ times the fiducial run over $t=26{,}410$--$26{,}700\,{\rm M}$. However, $\Phi_{\rm B,BH2}$ is larger by a factor of $\simeq1.7$ (Fig.~\ref{fig:bh2_resolution_diagnostics}b), indicating that the secondary-local magnetic flux and jet/outflow power remain resolution sensitive.

\begin{acknowledgments}
This research was supported by the National Key Research
and Development Program of China (grant No. 2023YFE0101200), the National Natural Science Foundation of China (grant No. 12273022 and 12511540053), and the Shanghai Municipality Orientation Program of Basic Research for International Scientists (grant No. 22JC1410600). 
ZY acknowledges partial support from a STFC Stephen Hawking Fellowship and a UKRI-STFC SA grant awarded to UCL-MSSL.
CMF is supported by the DFG research grant “Jet physics on horizon scales and beyond'' (Grant No. 443220636) within the DFG research unit “Relativistic Jets in Active Galaxies'' (FOR 5195). HXJ acknowledges support from the Tsung-Dao Lee PhD Student Fellowship Program.
The simulations were performed on the TDLI-Astro, Pi2.0, and Siyuan-I cluster at Shanghai Jiao Tong University. The authors gratefully acknowledge Mr. Shuoqi Wang for his kind assistance with the GPU facilities.  
\end{acknowledgments}

\begin{contribution}

\end{contribution}

%
\facilities{}

\software{AthenaK \citep{Stone2024,2025ApJS..278...50Z}, BHOSS \citep{2012A&A...545A..13Y,2020IAUS..342....9Y}, Python, NumPy, Matplotlib}

\bibliography{sample701}{}
\bibliographystyle{aasjournalv7}



\end{document}